\documentclass[twocolumn,twocolappendix]{aastex63}

\usepackage{xcolor}

\newcommand{\Mstar}{$M_{\ast}$}

\newcommand{\mgas}{$M_{\rm gas}$}
\newcommand{\lir}{$L_{\rm IR}$}
\newcommand{\md}{$M_{\rm dust}$}
\newcommand{\avu}{$\langle U \rangle$}
\newcommand{\msol}{M$_{\odot}$}
\newcommand{\fdust}{$f_{\rm dust}$}
\newcommand{\fgas}{$f_{\rm gas}$}
\newcommand{\td}{$T_{\rm d}$}
\newcommand{\h}{$Herschel$}
\newcommand{\zphot}{$z_{\rm phot}$}
\newcommand{\zspec}{$z_{\rm spec}$}
\received{February 26, 2021}
\revised{June 28, 2021}
\accepted{July 10, 2021}
\submitjournal{ApJ}

\shorttitle{ISM Evolution}
\shortauthors{Kokorev et al.}

\usepackage{amsmath}
\usepackage[normalem]{ulem}
\usepackage[para]{threeparttable}
\begin{document}

\title{The Evolving Interstellar Medium of  Star-Forming Galaxies, as traced by \texttt{Stardust}\footnote{\url{https://github.com/VasilyKokorev/stardust/}}}

\correspondingauthor{Vasily Kokorev}
\email{vasilii.kokorev@nbi.ku.dk}

\author{Vasily I. Kokorev}
\affiliation{Cosmic Dawn Center (DAWN), Jagtvej 128, DK2200 Copenhagen N, Denmark}
\affiliation{Niels Bohr Institute, University of Copenhagen, Blegdamsvej 17, DK2100 Copenhagen \O, Denmark}

\author{Georgios E. Magdis}
\affiliation{Cosmic Dawn Center (DAWN), Jagtvej 128, DK2200 Copenhagen N, Denmark}
\affiliation{DTU-Space, Technical University of Denmark, Elektrovej 327, DK2800 Kgs. Lyngby, Denmark}
\affiliation{Niels Bohr Institute, University of Copenhagen, Blegdamsvej 17, DK2100 Copenhagen \O, Denmark}

\author{Iary Davidzon}
\affiliation{Cosmic Dawn Center (DAWN), Jagtvej 128, DK2200 Copenhagen N, Denmark}
\affiliation{Niels Bohr Institute, University of Copenhagen, Blegdamsvej 17, DK2100 Copenhagen \O, Denmark}

\author{Gabriel Brammer}
\affiliation{Cosmic Dawn Center (DAWN), Jagtvej 128, DK2200 Copenhagen N, Denmark}
\affiliation{Niels Bohr Institute, University of Copenhagen, Blegdamsvej 17, DK2100 Copenhagen \O, Denmark}

\author{Francesco Valentino}
\affiliation{Cosmic Dawn Center (DAWN), Jagtvej 128, DK2200 Copenhagen N, Denmark}
\affiliation{Niels Bohr Institute, University of Copenhagen, Blegdamsvej 17, DK2100 Copenhagen \O, Denmark}

\author{Emanuele Daddi}
\affiliation{CEA, Irfu, DAp, AIM, Universit\`e Paris-Saclay, Universit\`e de Paris, CNRS, F-91191 Gif-sur-Yvette, France}

\author{Laure Ciesla}
\affiliation{Aix Marseille Univ, CNRS, CNES, LAM, Marseille, France}
\affiliation{CEA, Irfu, DAp, AIM, Universit\`e Paris-Saclay, Universit\`e de Paris, CNRS, F-91191 Gif-sur-Yvette, France}

\author{Daizhong Liu}
\affiliation{Max-Planck-Institut f\"ur Extraterrestrische Physik (MPE), Giessenbachstr. 1, D-85748 Garching, Germany}

\author{Shuowen Jin}
\affiliation{Instituto de Astrofísica de Canarias (IAC), E-38205 La Laguna, Tenerife, Spain}
\affiliation{Universidad de La Laguna, Dpto. Astrofísica, E-38206 La Laguna,Tenerife, Spain}

\author{Isabella Cortzen}
\affiliation{Institut de Radioastronomie Millim\'etrique (IRAM), 300 ruede la Piscine, 38400 Saint-Martin-d’H\'eres, France}

\author{Ivan Delvecchio}
\affiliation{INAF - Osservatorio Astronomico di Brera, via Brera 28, I-20121, Milano, Italy \& via Bianchi 46, I-23807, Merate, Italy}
\affiliation{CEA, Irfu, DAp, AIM, Universit\`e Paris-Saclay, Universit\`e de Paris, CNRS, F-91191 Gif-sur-Yvette, France}

\author{Clara Gim\'enez-Arteaga}
\affiliation{Cosmic Dawn Center (DAWN), Jagtvej 128, DK2200 Copenhagen N, Denmark}
\affiliation{Niels Bohr Institute, University of Copenhagen, Blegdamsvej 17, DK2100 Copenhagen \O, Denmark}

\author{Carlos G\'omez-Guijarro}
\affiliation{CEA, Irfu, DAp, AIM, Universit\`e Paris-Saclay, Universit\`e de Paris, CNRS, F-91191 Gif-sur-Yvette, France}

\author{Mark Sargent}
\affiliation{Astronomy  Centre,  Department  of  Physics  and  Astronomy,University of Sussex, Brighton, BN1 9QH, UK}

\author{Sune Toft}
\affiliation{Cosmic Dawn Center (DAWN), Jagtvej 128, DK2200 Copenhagen N, Denmark}
\affiliation{Niels Bohr Institute, University of Copenhagen, Blegdamsvej 17, DK2100 Copenhagen \O, Denmark}

\author{John R. Weaver}
\affiliation{Cosmic Dawn Center (DAWN), Jagtvej 128, DK2200 Copenhagen N, Denmark}
\affiliation{Niels Bohr Institute, University of Copenhagen, Blegdamsvej 17, DK2100 Copenhagen \O, Denmark}



\begin{abstract}
We analyse the far-infrared properties of $\sim$\,5,000 star-forming galaxies at $z<4.5$, drawn  from the deepest, super-deblended catalogues in the GOODS-N and COSMOS fields. We develop a novel panchromatic SED fitting algorithm, \texttt{Stardust}, that models the emission from stars, AGN, and infrared dust emission, without relying on energy balance assumptions. Our code provides robust estimates of the UV-optical and FIR physical parameters, such as the stellar mass ($M_*$), dust mass (\md), infrared luminosities (\lir) arising from AGN and star formation activity, and the average intensity of the interstellar radiation field (\avu).
Through a set of simulations we quantify the completeness of our data in terms of \md, \lir\ and \avu, and subsequently characterise the distribution and evolution of these parameters with redshift. We focus on the dust-to-stellar mass ratio (\fdust), which we parametrise as a function of cosmic age, stellar mass, and specific star formation rate. The \fdust\ is found to increase by a factor of 10 from $z=0$ to $z=2$ and appears to remain flat at higher$-z$, mirroring the evolution of the gas fraction. We also find a growing fraction of warm to cold dust with increasing distance from the main sequence, indicative of more intense interstellar radiation fields, higher star formation efficiencies and more compact star forming regions for starburst galaxies. Finally, we construct the dust mass functions (DMF) of star-forming galaxies up to $z=1$ by transforming the stellar mass function to DMF through the scaling relations derived here. The evolution of \fdust\ and the recovered DMFs are in good agreement with the theoretical predictions of the Horizon-AGN and IllustrisTNG simulations.

\end{abstract}



\section{Introduction} \label{sec:intro}
One of the main mechanisms driving galaxy evolution is the interaction between the interstellar medium (ISM), primarily consisting of gas and dust, and the radiation field induced by stellar activity. 
In this context, dust poses challenges in the detection of the  UV/optical emission of galaxies and in the interpretation of these observations in terms of physical properties (e.g.\ star formation rate (SFR), stellar mass \Mstar, etc), but also is an important tracer of star formation and ISM in the FIR observations.
At the same time, dust shields cold molecular hydrogen from ionising photons, and facilitates the collapse of molecular gas and subsequent star formation \citep{goldsmith01,krumholz11,narayanan11,narayanan12,narayanandave12}. As such, dust plays a critical role in the life cycle of galaxies and offers observational signatures regarding their past evolutionary stages. 

The impressive variety of infrared and millimetre facilities commissioned in the last few decades  have propelled the extragalactic ISM studies at an ever increasing number, redshift, and detail. Indeed, the enormous observational efforts manifested by the large far-IR/mm imaging and spectroscopic surveys such as PEP \citep{lutz2011}, GOODS-Herschel \citep{elbaz2011}, PHIBBS \citep{tacconi13}, S2CLS \citep{geach2017} and many others (e.g.\  \citealt{oliver12,magnelli2013,walter16,saintonge17,dunlop17,maddox18,franco20,valentino20,bethermin20,reiter20}, for a review see 
\citealt{carilli13} and \citealt{hodge20}) have yielded a wealth of multi-wavelength data sets and have advanced our understanding of galaxy evolution through scaling relations that have been used to guide simulations and theoretical models (e.g.\ \citealt{dekel09,popping14,narayanan15,lagos15,lagos20,popping17,dave17,dave20}).

In the evolutionary picture that is emerging from the analysis of the FIR/mm surveys, the majority of star-forming galaxies (SFGs) follow a tight relation -- the Main Sequence (MS) -- between the SFR and \Mstar\ with an increasing normalisation factor (specific star-formation rate, sSFR= SFR/\Mstar) at least up to $z \sim 4$  \citep{brinchmann04,daddi07,elbaz07,noeske2007,salim07,chen09,pannella09,santini09,daddi10,elbaz2010,oliver10,magdis10,elbaz2011,karim11,rodighiero11,shim11,lee12,reddy12,salmi12,whitaker12,zahid12,kashino13,moustakas13,rodighiero14,sargent14,steinhardt14,sobral14,speagle14,whitaker14,lee15,schreiber15,shivaei15,tasca15,whitaker15,erfanianfar16,kurczynski16,santini17,pearson18,leslie20}. This  elevation of sSFR with lookback-time, which broadly mirrors the overall increase of the star-formation rate density of the Universe from $z=0$ to $z=2-3$, is also followed by a similar rise in the gas fraction (\fgas\ = \mgas/\Mstar) of SFGs (e.g.\  \citealt{daddi10,geach11,magdis2012,magdis12b,tacconi13,liu19a,liu19b}).
Nevertheless, for fixed \Mstar, the increase in star-formation efficiencies (SFE=SFR/\mgas) surpasses that in \mgas, resulting in higher SFR, an activity that coupled with the observed \Mstar -- size evolution (e.g.\ \citealt{vanderwel14}) instils warmer luminosity-weighted dust temperatures of the ISM as a function of redshift (e.g.\ \citealt{magdis2012,magdis17,magnelli2014,bethermin15,casey18,schreiber18,liang19,cortzen20}).

Although the purity of the MS as a measure of the evolutionary stage of a galaxy has recently been challenged (e.g.\ \citealt{elbaz18,puglisi19,valentino20}), it appears that the majority of SFGs grow along the MS by secularly converting (and hence depleting) their gas mass reservoirs into stars (e.g.\ \citealt{daddi10,dave12,lilly13,tacchella16}), with a high degree of uniformity in the properties of their ISM (at fixed redshift). On the other hand, galaxies above the MS (starbursts; hereafter SBs) are primarily characterised by elevated sSFR, SFE and dust temperature ($T_{\rm d}$) with respect to the average star-forming population at their corresponding redshift (e.g.\ \citealt{daddi10,magdis2012,magnelli2014,scov17,tacconi18,silverman18}). Galaxies below the MS are mainly post starbursts or quenched systems with low levels of star-formation activity, low gas fractions and cold \td\ (e.g.\ \citealt{williams20,magdis21}).

At the core of the aforementioned results is the robustness of the derived FIR properties, namely the total infrared luminosity ($L_{\rm IR}$), the dust mass ($M_{\rm dust}$), the mean intensity of the radiation field ($\langle U \rangle$ $\propto$ $L_{\rm IR}/M_{\rm dust}$) and $T_{\rm d}$. These quantities and their emerging evolution with redshift rely on the availability of FIR/mm data and on selection effects. In this regard, while the ongoing ALMA observations are  quickly filling the gap in  resolution and sensitivity  between the available UV/optical/NIR (sub-arcsecond) data and the coarse resolution of the confusion-limited SCUBA2 and SPIRE/\h\ surveys, the vast majority of the SFG samples with available FIR photometry are still restricted to the latter. Thus, FIR studies are still largely focusing either on the FIR luminous and most massive SFGs, on limited and possibly non-homogeneous or biased ALMA samples, or on stacking techniques.  

Moreover, the derived measurements of \md\ and \td\ heavily rely on the adopted models and fitting techniques (e.g.\ \citealt{dale14,berta16}). Indeed, without a coherent and homogeneous treatment of the data sets it is impossible to overcome systematic effects that could distort the observed trends. On top of that, recent high-resolution observations  with ALMA indicate that the UV/optical and millimetre emission of some of high-z SFGs are spatially distinct (e.g.\ \citealt{hodge16}), posing challenges to the widely adopted energy balance assumption that is inherent in most multi-wavelength fitting codes. Similarly, there is an ever increasing number of IR bright yet optically faint/dark sources (e.g.\ \citealt{wang16,jin19,casey19,franco20}) that an energy balance approach would have technical difficulties to accommodate. 

Finally, while many studies have focused on \fgas, the evolution of the dust fraction (\fdust\ = \md/\Mstar) and the dust mass function (DMF) with redshift have not been scrutinised to the same extent (\citealt{dunne11,magdis2012,santini2014,tan14,bethermin15,driver18,magnelli20,donevski20}). Given that the use of \md\ as a proxy of \mgas\ either through the metallicity dependent dust-to-gas mass ratio technique (e.g.\ \citealt{leroy2011,eales12,magdis2011,magdis2012,berta16,tacconi18}) or (indirectly) through the monochromatic flux densities in the Rayleigh-Jeans (R-J) tail of the SED (e.g.\ \citealt{groves15,scov17}) has gained momentum, a proper investigation of the evolution of \fdust\ and DMF with redshift is necessary, and remains to be done. More importantly, the \fdust, the DMFs and in general the life cycle of dust are key in our understanding of metal enrichment processes and dust production mechanisms. These derived properties are also critical parameters of semi-analytical and analytical models that  couple the evolution of stars, metals, and gas  \citep{lacey16,popping17,imara18,cousin19,vijayan19,lagos19,pantoni19}, as well as of cosmological simulations that also trace the dark matter component \citep{hayward13,narayanan15,mckinnon17,dave19,aoyama19}.
 
These considerations provide motivation for a coherent and homogeneous analysis of the full population of IR galaxies that are detected in the recently constructed, state-of-the-art `Super-Deblended' FIR catalogues in two of the most extensively studied cosmological fields,  the Great Observatories Origins Deep Survey North \citep[GOODS-N;][]{goods}  and the Cosmic Evolution Survey \citep[COSMOS;][]{cosmos}. These catalogues are built using the `Super-Deblending' technique \citep{sdc1,sdc2} that allows prior-based source extraction from highly confused \emph{Herschel} and \emph{SCUBA+AzTEC} maps, yielding robust UV to radio photometry for thousands of individually detected galaxies. To model the observed SEDs we built and make publicly available a novel, time efficient and panchromatic SED fitting algorithm that we use to infer and explore the evolution and the variations of IR properties of SFGs (\md, \td, \avu, \fdust, \fgas, DMF) out to $z\sim4$ and compare those to recent theoretical predictions. The catalogues with the derived FIR parameters for the full sample are also publicly released. 

The paper is organised as follows. In \autoref{sec:sample} we describe the data sets used in this work, while \autoref{sec:sed} introduces our SED fitting algorithm. In \autoref{sec:sims} we perform various simulations to determine the robustness of our sample, as well as the limiting \md. \autoref{sec:genprop} presents our physical estimates for each galaxy in the sample, and their evolution with $z$. In \autoref{sec:dms} we analyse the evolution of \fdust\ and calculate the DMF through the conversion of the stellar mass function (SMF).
In \autoref{sec:gas} we constrain the evolution of \fgas\ and compare it to the literature results. We discuss the implications that our findings have in \autoref{sec:disc}, and present our main conclusions and summary in \autoref{sec:conc}.

\par Throughout this paper we assume a flat $\Lambda$CDM cosmology with $\Omega_{\mathrm{m},0}=0.3$, $\Omega_{\mathrm{\Lambda},0}=0.7$ and H$_0=70$ km s$^{-1}$ Mpc$^{-1}$, and a \citet{chabrier} initial mass function (IMF).

\section{Panchromatic catalogues and sample selection} \label{sec:sample}

\subsection{GOODS-N `Super-Deblended' Catalogue}
We first consider the `Super-Deblended' photometric catalogue (hereafter SDC1) constructed by \citet[][]{sdc1} using the  FIR and sub-mm images in GOODS-N. These images come from the \textit{Herschel} Space Observatory (PACS and SPIRE instruments, see \citealt{elbaz2011,magnelli2013}) and the  ground-based  facilities SCUBA-2 (850\,$\mu$m; \citealt{geach2017}) and AzTEC+MAMBO (1.16\,mm; \citealt{penner11}).

Several novelties introduced in SDC1 are particularly useful for our analysis. First, detections from deep Spitzer IRAC and VLA 20 cm \citep{owen18} are used as a prior for the positions of the blended FIR/sub-mm sources. Second, the SED information from shorter-wavelength photometry is also used as a prior for subtracting lower redshift sources. This substantially decreases blending degeneracies and allows for a robust photometry extraction of sources at longer wavelengths. Moreover, the authors estimated more realistic photometry uncertainties for each photometric measurement with extensive Monte Carlo simulations. These improvements allow for deeper detection limits and statistically reliable estimates (both measurements and uncertainties) in the FIR+mm bands.\par
The SDC1  contains 3,306 priors in total, including over 1,000 FIR+mm detections. All sources have photometric redshifts and stellar masses inferred by \texttt{EAZY} \citep{brammer08}, and \texttt{FAST} \citep{kriek09} respectively, based on the 3D-HST UV-near-IR \citep{Skelton14} and \citealt{pannella2015} GOODS-N catalogues. Following \citet[][]{sdc1}, we extend the photometric coverage of the published SDC1 catalogue to shorter wavelengths by cross-matching with the 3D-HST UV-near-IR \citep{Skelton14} and \citealt{pannella2015} GOODS-N catalogues. Approximately half of the objects within the catalogue are spectroscopically confirmed. However, as mentioned by the authors, the outer perimeter of the GOODS-N area contains objects with high instrumental noise in the $24\,\mu$m prior image that may impair the extraction process. We therefore choose to limit our analysis to the central 134\,arcmin$^{2}$  with reliable photometry (flag \texttt{goodArea} = 1 in SDC1). This reduced our final sample to 2,344 objects. 

\begin{figure}
\begin{center}
\includegraphics[width=0.5\textwidth]{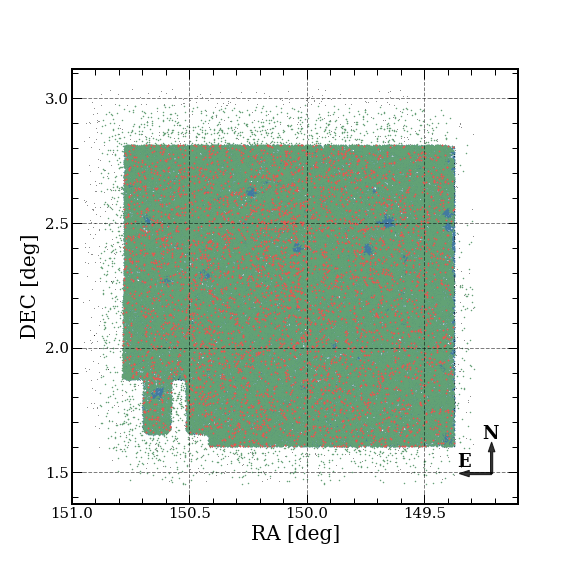}
\caption{COSMOS sky map. The blue regions and black points represent sources from \citet{muzzin13} with \texttt{goodArea} = 1 and 0 respectively (according to the quality flag in \citealp{sdc2} catalogue). The \citet{laigle2016} coverage is shown in green. Sources that we use for our analysis are indicated in red.\\
}
\label{fig:cosm_area}
\end{center}
\end{figure}

\begin{figure*}
\begin{center}
\includegraphics[width=0.95\textwidth]{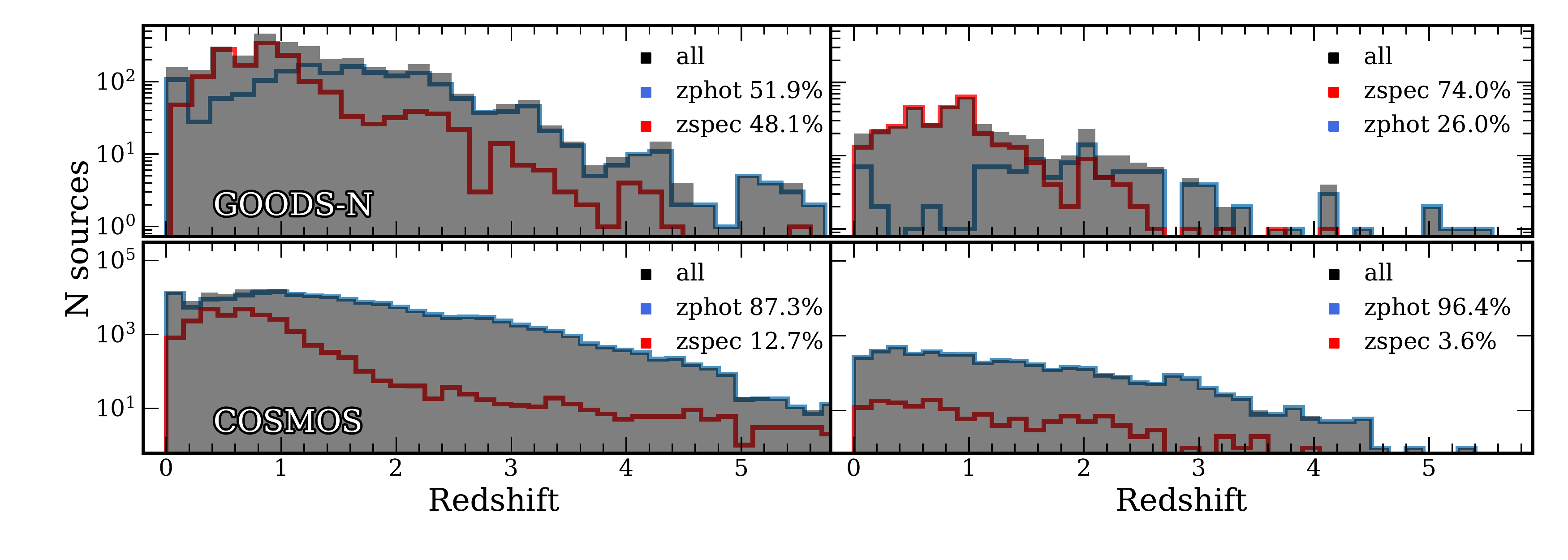}
\caption{Redshift distributions of the sources considered in the present work. Both the photometric and spectroscopic redshifts were taken from the corresponding SDCs. The left and right panels show the redshift distribution of the original, full catalogue and of the final sample that meets the selection criteria described in \autoref{sec:selection}, respectively.}
\label{fig:goods-z}
\end{center}
\end{figure*}

\subsection{COSMOS `Super-Deblended' Catalogue}
We supplement our study with the `Super-Deblended'  catalogue (SDC2 hereafter) presented in \citet{sdc2}. The catalogue covers  1.7\,deg$^2$ in COSMOS, in the same bands as in SDC1, plus additional MAMBO data at 1.2\,mm \citep{bertoldi2007}.\par
In practice, the deblending methodology remains identical to that adopted in SDC1, with one primary difference: an additional step selecting a highly complete sample of priors in $K_{s}$ band from the UltraVista catalogues \citep{uvista}. The resulting 24\,$\mu$m detections are then combined with the mass-limited sample of $K_{s}$ sources in order to fit the remaining bands in the catalogue. \par

The final input dataset contains 195,107 priors, with 13\% of them having spectroscopic confirmation. The authors highlight that only 11,220 objects are in fact detected over the 100-1200\,$\mu$m range.
Similarly to GOODS-N, we impose the \texttt{goodArea = 1} flag to only include sources that are present in the UltraVista Data Release 4 area (\citealt{uvista}). 

 We note that for their catalogue, \citet{sdc2} used a combination of \citet{laigle2016} and \citet{muzzin13} (M13 hereafter) catalogues. The M13 catalogue has an advantage in that it does not completely remove the sources around saturated stars, which has a positive effect on the number counts, however the reduced quality of the photometry could lead to unreliable estimates for photometric redshifts ($z_\mathrm{phot}$), as well as any parameters extracted by fitting optical templates. To be on the safe side, our analysis of SDC2 will only focus on the \citealt{laigle2016} sources, which narrows down the number of objects in the input catalogue to 186,549 and the total area to 1.38 deg$^2$. The COSMOS 2015 catalogue \citep{laigle2016}, also comes with a plethora of UV-optical photometry, spanning an additional $\sim 20$ bands, as well as photometric redshifts and stellar mass estimates by \texttt{LePhare} \citep{lephare1,lephare2}. We exploit these data by cross-matching the same COSMOS 2015 UV-optical photometry that was used to derive $M_*$ and redshift to SDC2, thus extending our photometric coverage.
 In total, the merged catalogues consist in $\sim 40$ bands.
 
 For posterity, in  \autoref{fig:cosm_area}  we  present  the  UltraVista Data Release 2 map \citep{laigle2016,davidzon2017} (L16 area), with the regions where the star subtraction took place being clearly identified. In  addition  to  that, on the West border of the survey there exists a number of sources falling outside of the L16 area. These only have UltraVista coverage and lack additional \textit{Subaru} photometry, which could affect the reliability of the $z_\mathrm{phot}$.

\subsection{Sample Selection}
\label{sec:selection}
The primary parameters that we can derive from observing the rest-frame FIR emission are the total infrared luminosity (\lir), the dust mass (\md), and $T_{\rm d}$ or equivalently the intensity of interstellar radiation field (\avu) in the \citet{dl07} dust model. To obtain robust estimates for these parameters, an adequate multi-wavelength sampling of a galaxy's SED is required. As such, constraining the IR-peak is necessary for a robust \lir\ estimate, while detections in the long wavelength regime (Rayleigh-Jeans) are imperative to capture the emission from cold dust.

With these considerations in mind, after the initial cleaning of the catalogues described above, we perform our sample selection based on the following requirements:

\begin{itemize}
    \item Detection at a S/N$ > 3$ significance in at least three FIR to sub-mm bands from 100\,$\mu$m to 1.2\,mm \footnote{Three bands are also required to reduce fitting degeneracies}.
    \item Available \zphot\ (or \zspec) and \Mstar\ estimates inferred by UV to near-IR photometry
\end{itemize}

After the quality cuts and the selection criteria, we are left with 4,331 objects in COSMOS and 585 sources in GOODS-N, which constitute our sample. We have additionally identified 75 objects within SDC2, that fulfil our FIR detection criteria, but despite having a well sampled UV-optical photometry, do not have either $z_{\rm phot}$ or $M_*$ estimates. We subsequently fit these sources with \texttt{EAZY} \citep{brammer08}, extract their $z_{\rm phot}$, as well as other UV-optical properties, and add them to our final catalogue. This brings the total amount of COSMOS sources to 4,406.
The \zphot\ and \zspec\ distributions of the final sample in the two fields are presented in \autoref{fig:goods-z} (right).
As we will discuss later, a third criterion requesting at least one detection at $\lambda_{\rm rest} > 150\,\mu$m will be imposed in order to define a sub-sample with robust \md\ estimates.

\section{SED fitting}
\label{sec:sed}
\subsection{The Inventory of Available SED fitting routines}
Prior to providing the description of our methodology, we believe it is important to outline and present a brief introduction of the available SED fitting codes that deal with a three component fitting approach, namely combining the optical, AGN and dust emissions. These include the well established and widely used energy-balance routines such as \texttt{CIGALE} \citep{cigale1,cigale2,cigale3}, \texttt{MAGPHYS} \citep{dacunha08,magphys} as well as its AGN template extension presented in \citealt{chang17}. These have inspired more novel and sophisticated approaches that optimise the template libraries to achieve significant improvements in computational speeds - \texttt{SED3FIT} \citep{berta13}, or adopt MCMC methods when extracting best fit parameters such as \texttt{AGNfitter} \citep{agnfitter} and \texttt{Prospector-$\alpha$} \citep{leja18}. These efforts are not just limited to published software packages, with many authors implementing their own routines for a panchromatic model analysis (e.g. see \citealt{feltre13,symeonidis13}).
\subsection{Basic Description of the \texttt{Stardust} Fitting Code}

\begin{figure*}
\begin{center}
\includegraphics[width=0.75\textwidth]{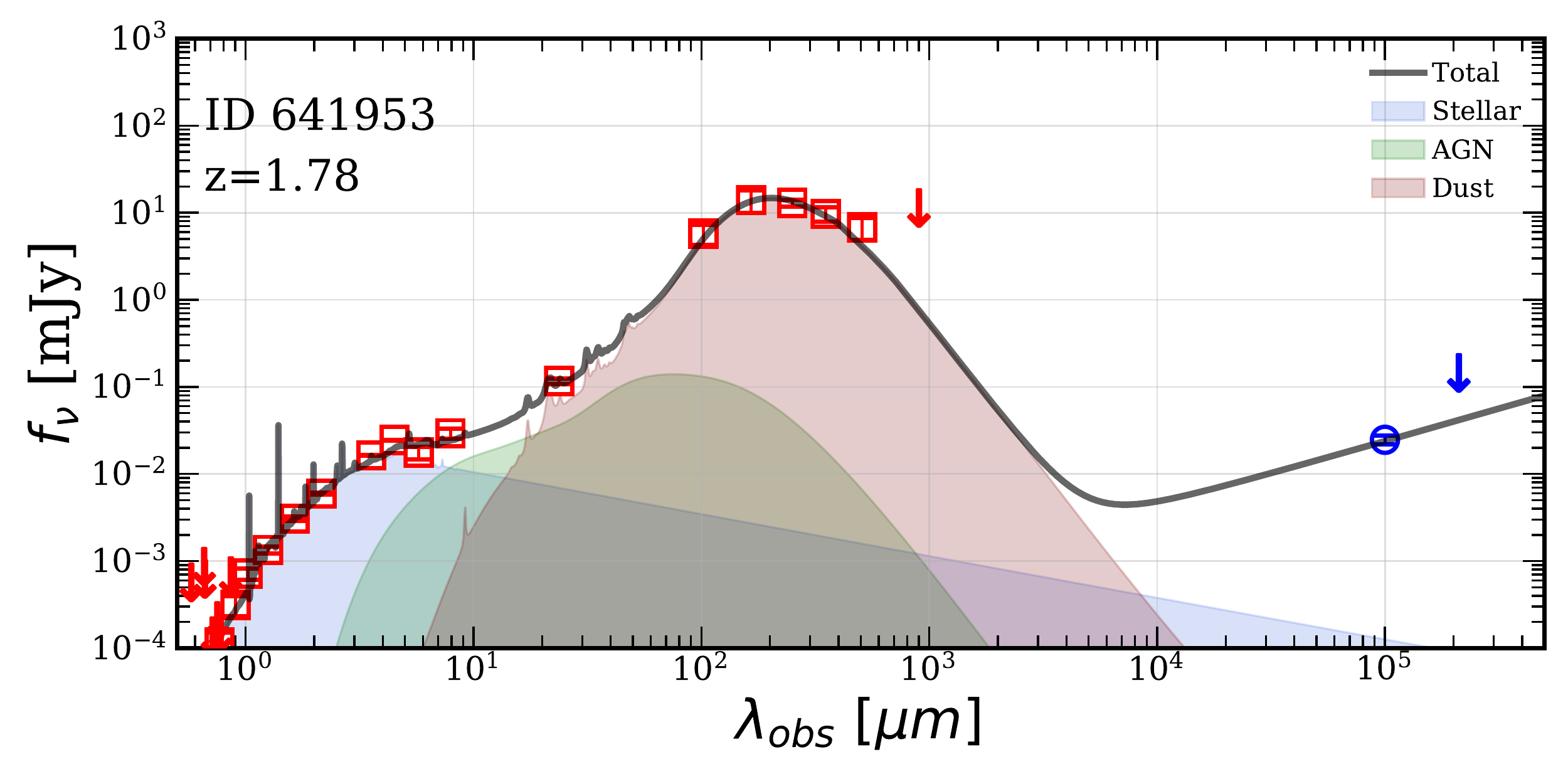}
\includegraphics[width=0.95\textwidth]{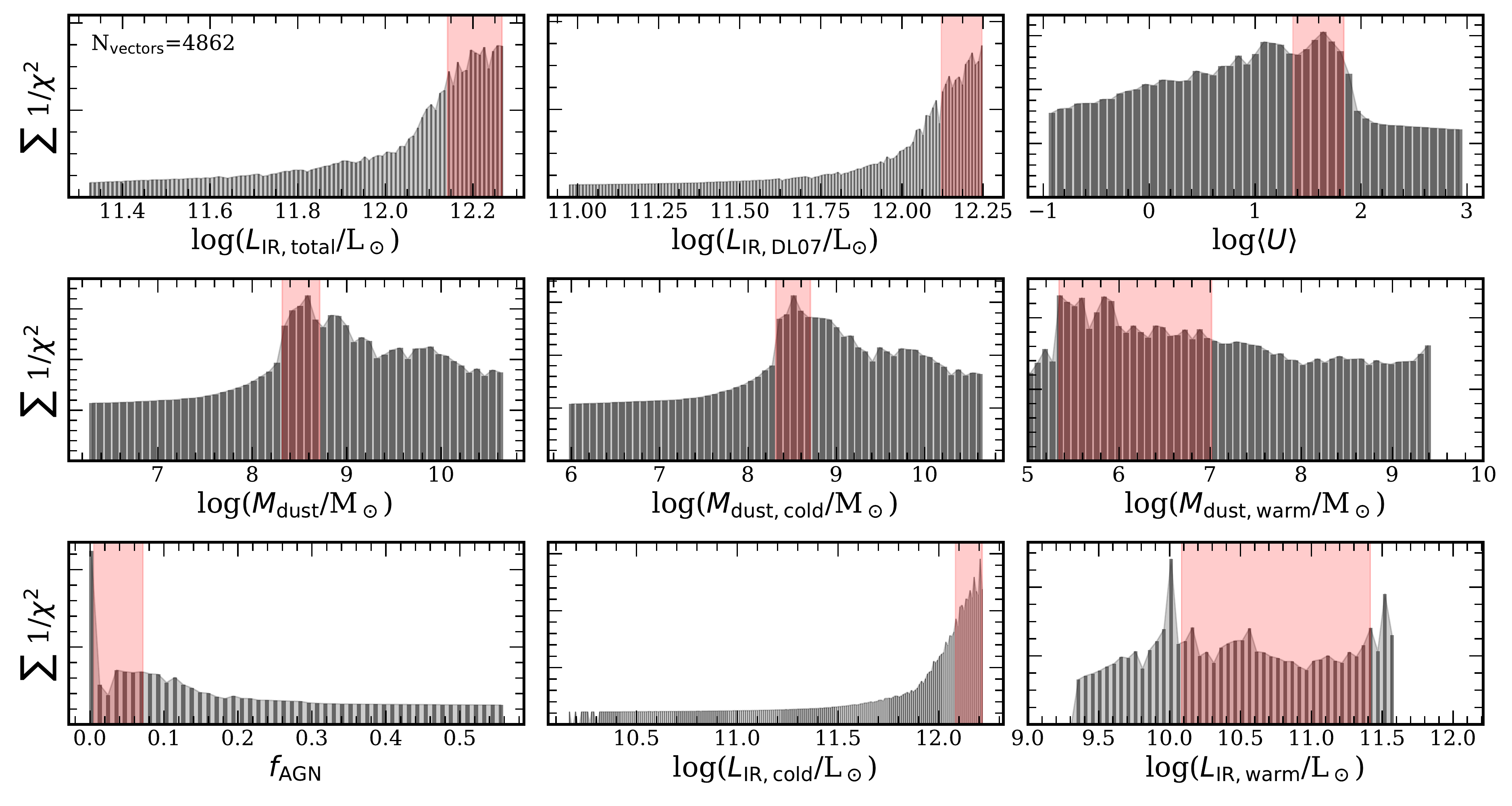}
\caption{\textbf{Top:} Example of an observed and best fit SED, as obtained with \texttt{Stardust} code for a \zphot\ = 1.78 galaxy (ID641953) drawn from the SDC2 sample. The squares and circles represent the S/N $> 3$ photometric detections while $3\sigma$ upper limits are shown as downward arrows. Red symbols represent the SDC2 photometry that was used in our fitting routine, while blue symbols show the radio measurements at 1.4\,GHz and 3\,GHz that were not included in the fit. Instead, the radio part of the model SED is based on the \lir\ - $L_{\rm 1.4\,GHz}$ relation of \citet{delv20}. The shaded red, green and blue regions, correspond to the dust, AGN and stellar components respectively. A linear combination of all three is given by a solid black line. \textbf{Bottom:} The $\chi^{2}$ distributions for the main derived parameters. The shaded red areas enclose solutions for which $\Delta\chi_{\nu}^{2}= 1$.}
\label{fig:seds}
\end{center}
\end{figure*}

To model the extensive photometric coverage of the galaxies in our sample we develop a new, panchromatic SED fitting tool: \texttt{Stardust}. The code performs a multi-component fit that linearly combines stellar libraries with AGN torus templates and IR models of dust emission arising from star-formation (SF-IR). This approach, which is very similar to that presented in \citep{liu21},  has a number of key differences compared to the currently existing SED fitting codes such as \texttt{MAGPHYS}, \texttt{CIGALE} and \texttt{SED3FIT}.
\par
First, the three components (stellar, AGN and SF-IR) are fit simultaneously yet independently from each other, without assuming an energy balance between the absorbed UV/optical radiation and the IR emission. The energy balance approach relies on the assumption that fitted stellar and dust emissions are co-spatial, i.e.\ the process of UV absorption and subsequent re-emission at IR wavelength happen in the same environment (\citealt[][e.g. see their Section 2.1 and 2.2]{dacunha08}). However, the detected stellar and dust distributions within a galaxy are not always physically connected.  Resolved observations of high-$z$ dusty-SFGs (DSFGs) \citep{simpson15,cgg18,franco18,hodge16,hodge19,kaasinen20} have revealed spatial offsets between the extent of the dust and stellar emitting sizes of high$-z$ DSFGs, with the former being on average more compact \citep[e.g.][]{chen17,tadaki17,calistrorivera18,cochrane21}. While energy balance is anticipated to apply universally, the aperture and sensitivity limitations of our observations cast a concern on how reliably we can bring these components together in the same panchromatic fit. These observations are also theoretically supported by radiative transfer codes \citep[e.g. SKIRT,][]{cohrane19} and hydrodynamical simulations \citep[e.g. IllustrisTNG,][]{pillepich19,popping21}.

Moreover, the detection of `HST-dark' galaxies, i.e. sources that are undetected in the UV/optical bands, and thus do not have the photometry to constrain neither dust absorption nor stellar emission, but are bright in the IRAC and FIR/mm bands  \citep{wang16,franco18}, pose another technical challenge to the correct application of the energy balance approach. Finally, the manner in which dust and stellar emission are connected, by assuming a single dust attenuation law and dust composition, can have a significant impact on derived parameters, as these recipes have been shown not to apply universally \citep{buat19}. In summary, while the premise of the energy balance routines is undoubtedly elegant and in most cases physically motivated, we choose to use independent stellar, AGN, and dust components to better focus on the dust properties themselves. \par
Furthering this complex picture, it is important to note the presence of cold diffuse dust, that is being heated by an older stellar population, rather than an ongoing star-formation activity, (see \citealt{boquien11,bendo12,galametz14,hayward14}).
However, when dealing with non-resolved observations, the diffuse dust and RJ-tail emissions are highly degenerate, and as such none of the aforementioned codes, nor \texttt{Stardust}, utilise these templates.

The other advantage of \texttt{Stardust} is related to the $\chi^{2}$ minimisation approach to select the best fit models. Instead of finding the solution from a vast library of  pre-compiled templates, we devise an optimisation method  to find the best linear combination of a much smaller set of `basic' templates (similar to eigenvectors in principal component analysis). This is the same approach adopted in the photometric redshift fitting code \texttt{EAZY} \citep{brammer08}. In our case, the basic templates are divided in three classes and the linear combination includes a sum of templates from these classes. The models used to create these templates are the following:

\begin{itemize}
\item Stellar library. We incorporate an updated version of the Stellar Population Synthesis (SPS) models described in \citet{brammer08}.
Although UV and optical photometry is not always available, the inclusion of the stellar component in the code is important in the NIR regime. In particular, it allows to better constrain the AGN contribution  (see \autoref{fig:stelvsnostel} and \autoref{sec:appendix}). This stellar library represents an optimised basis-set, where the non-negative linear combinations
of models can be considered to be the ``principal components'' of a much larger parent template catalogue (see \citealt{brammer08} and \citealt{br07}).

\item AGN library. We adopt empirically derived templates from \citet{Mullaney2011} describing AGN intrinsic emission from 6 to 100\,$\mu$m \footnote{Note: These templates do not account for X-ray selected QSOs. The flexible nature of the code however allows these templates to be manually injected if necessary.}. We include both  high- and low-luminosity templates (total of 2). Since these can be linearly combined, we do not include the median luminosity AGN template.
\item Infrared library. It consists in 4,862 \citealt{dl07} (DL07 hereafter) templates, with the additional updates from \citealt{draine14} (also see \citealt{aniano20}). These models describe the contribution from warm dust and polycyclic aromatic hydrocarbon (PAH) features in the photo-dissociation regions (PDR), together with cold dust in the diffuse part of the ISM. We consider\footnote{The modular structure of the code allows the user to decide which DL07 templates to use.} a wide array of values for the minimum radiation field ($U_{\rm min}$)  in the 0.1 $<$ $U_{\rm min}$ $<$ 50 range, as well as the fraction of the total dust mass locked in PAHs ($q_{\rm PAH}$) between 0 and 10 \%. We have fixed $U_{\rm max}=10^6$ and $\alpha=2$, as described in \citealt{magdis2012}. These templates are not linearly combined within their class, the algorithm instead chooses a single best-fit DL07 template.
\item Radio continuum. Data points in radio are not considered by our fitting routine, however they can be used a posteriori to quantify possible radio excess and further confirm the presence of an AGN, if needed. Our radio model is based on the radio-FIR correlation, described in \citet{delv20}, with a spectral index of $-0.75$. 
\end{itemize}

More details on the characteristics of the templates and the motivation for selecting them are provided in \autoref{sec:appendix}. With such a configuration, fitting a single object (including the computation of the uncertainties) with \texttt{Stardust} takes less than $10$ seconds\footnote{Tested on a i9-8950HK CPU.}, i.e.\ a factor of 8-10 faster than software like \texttt{CIGALE} (see \autoref{sec:appendix-d}), based on large pre-compiled template sets. If we choose to pre-compile all of our templates, instead of linearly combining them, the resulting model library would contain millions of possible SEDs, with computation time increasing by a significant amount. The code is also highly parallelised, which allows it to run on multiple threads simultaneously, thus achieving significant computation speeds on modern CPUs.

\subsection{Configuration of the Code}
For each object, the input consists of measured flux densities, their corresponding uncertainties and a redshift estimate.\footnote{It is also possible to manually define a rectangular filter at a desired wavelength, for cases where the filter transmission curve is not easily obtainable, e.g. ALMA.} The user must then choose the corresponding filter curves from the pre-compiled set, or upload their own. The individual template components can be switched off and on as an additional user input. The algorithm then outputs the best-fit FIR as well as AGN properties of the source. If the photometry is available, the UV-optical parameters are also produced. These can be summarised as follows:
\begin{itemize}
    \item The total infrared luminosity integrated over the SF-IR+AGN templates ($L_{\rm IR,total}$)\footnote{In this work we use terms $L_{\rm IR}$ and $L_{\rm IR,total}$ interchangeably.}, the total infrared luminosity associated with star-formation ($L_{\rm IR,DL07}$) and the relative contribution of the PDR component ($f_{\rm PDR}$) to $L_{\rm IR,DL07}$.
    \item The bolometric IR luminosity of the AGN ($L_{\rm AGN}$) and its fractional contribution ($f_{\rm AGN}$) to the total IR energy budget\footnote{The quality of the photometry in this work does not allow us to distinguish bolometric AGN contributions below $\sim 0.5\%$, and thus the non-zero entries below that threshold are treated here as zero.}.
    \item The total dust mass ($M_{\rm dust}$), the warm dust mass component heated by PDRs ($M^{\rm warm}_{\rm dust}$), the fraction of the total dust mass heated by PDRs ($\gamma$), the cold dust mass component ($M^{\rm cold}_{\rm dust}$) in the diffuse ISM,  and the fraction of the total dust mass locked in PAHs $q_{\rm PAH}$.
     \item The intensity of the radiation field at which the diffuse ISM is exposed to ($U_{\rm min}$), which is a proxy of the mass-weighted \td\ of the ISM,  and the mean radiation field intensity (\avu), which is a proxy of the luminosity-weighted \td. 
     \item The stellar mass ($M_*$), star-formation history (SFH), $E(B-V)$ and the unobscured SFR, if there is available optical photometry.
\end{itemize}
\par
\autoref{fig:seds} presents an example fit to one of the COSMOS galaxies, chosen for its prominent AGN contribution. The top panel of the Figure shows the data points and the best-fit model, with different colours for the four components listed above;  the bottom panel displays the $\chi^{2}$ distributions of all relevant IR quantities.

\begin{figure*}
\begin{center}
\includegraphics[width=.8\textwidth]{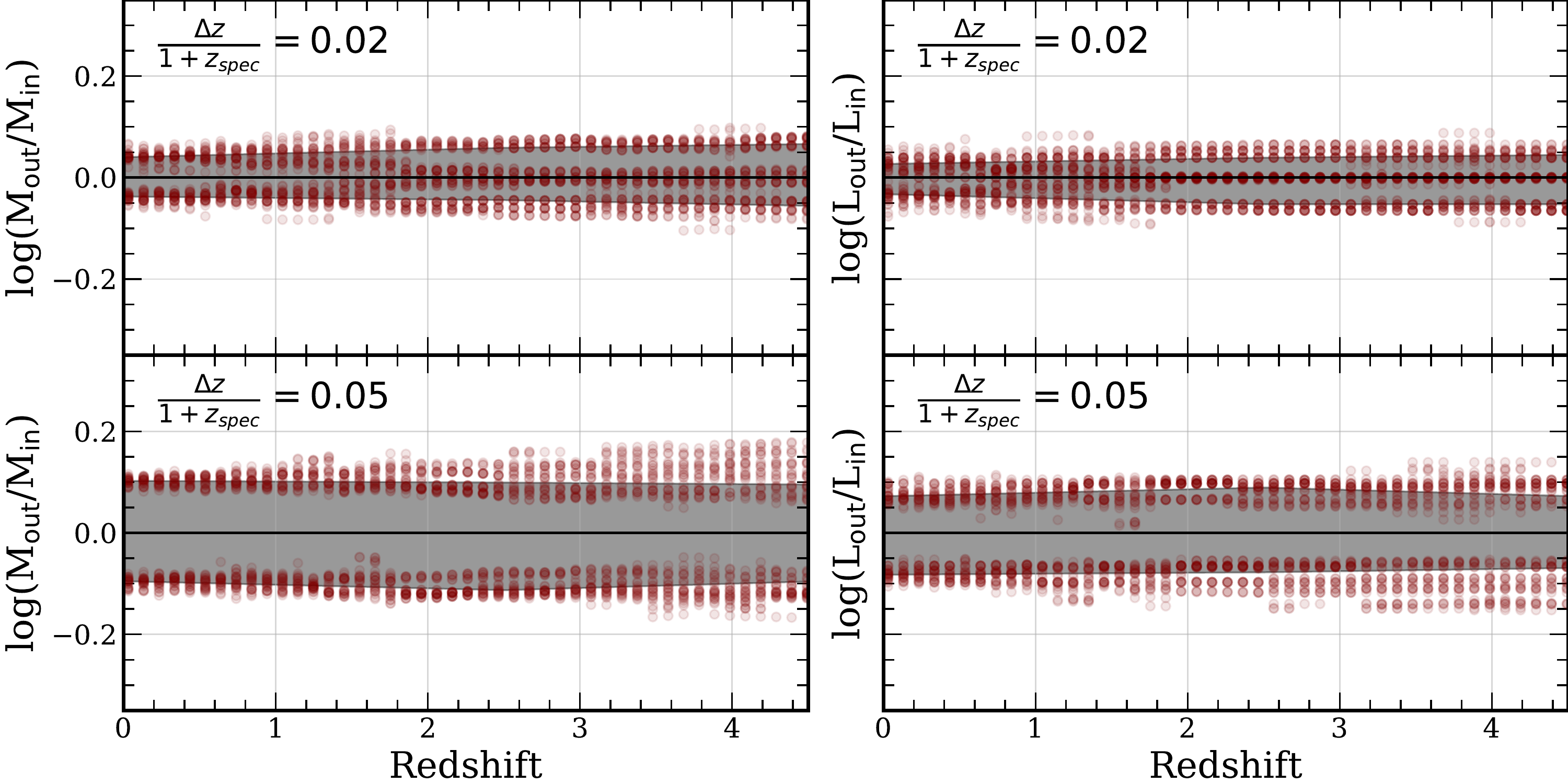}
\caption{Effect of the \zphot\ uncertainty in the derivation of \md\ (left panel) and \lir\ (right panel) assuming $\epsilon = 0.02$ (top) and $\epsilon = 0.05$ (bottom). Red circles represent the ratio of the output to input quantities from our simulations, as inferred by shifting the fitted redshift by $\pm$$\rm\Delta$$z$. The shaded regions cover the $68 \%$ confidence interval and the solid black line indicates a ratio equal to unity.}
\label{fig:photoz}
\end{center}
\end{figure*}

\subsection{Derivation of Uncertainties}

In order to estimate the errors associated with the derived quantities during the fitting procedure of  \texttt{Stardust}, we consider two main sources of uncertainty;  one concerning the linear combination coefficients of the best-fit optimisation and one linked to the broad-band photometric data. 

To quantify the linear combination uncertainty, we re-sample the best solution coefficients. A covariance matrix is first created by considering all of the templates that went into the best-fit solution. We draw the coefficients from a multivariate normal distribution whose median are given by the coefficient of the best-solution vector and the standard deviation is computed from the diagonalised covariance matrix. This is done $10^{4}$ times  to provide a good balance between robustness of the error estimates and computational speed. We recompute all the relevant FIR properties for each realisation of our routine. From resultant distributions we then define our lower and upper uncertainty as the 16th and 84th percentile confidence interval respectively. However, given the fact that only a single solution with a single coefficient is considered in the IR, the final uncertainty on the FIR properties is underestimated.
\par

We therefore also consider the observational uncertainty, that is primarily driven by the quality of the photometric data. We compute it by considering all possible solutions from our template library which fall within the $68 \%$ confidence interval range of the best-fit. This would correspond to a region where the solutions fall within $\Delta\chi_\nu^{2}= 1$, where $\Delta\chi_\nu^{2}=\chi_{\nu,i}^2-\chi_{\nu,\mathrm{best}}^2$, since the non-diagonal terms of the template covariance matrix are not zero \footnote{See sections 15.1 and 15.6 of \citet{press86} and \citet{avni}. Note that $\Delta\chi_\nu^{2}= 1$ only applies when marginalised over all other parameters.}.
We show these as red shaded areas on the bottom panel of \autoref{fig:seds}. The observational uncertainty is then derived as simply the width of the shaded region. \par
The final errors are computed as a quadrature sum of the linear combination uncertainty and the observational uncertainty.

\subsection{The Effect of Photometric Redshift Uncertainty}
To explore and quantify how the uncertainty in \zphot\ propagates into the estimates of $L_{\rm IR}$ and $M_{\rm dust}$, we built mock IR SEDs of 1,200 galaxies utilising a suite of 0.1$ < U_{\rm min} < 50$ DL07 models and place them uniformly in the $0.03 < z_{\rm true} < 5$ redshift range. Thus, each mock SED is characterised by a set of pre-defined $L_{\rm IR,in}$, $M_{\rm dust,in}$,  and $z_{\rm true}$ values. We then infer synthetic broad-band photometry in all IR bands available in \emph{SDC2} (24-1100$\,\mu$m) for each simulated galaxy. Since we are interested in the effect of the \zphot\ uncertainty on the derived FIR properties, to minimize any other possible sources of error (e.g. photometric uncertainty, poor photometric sampling of the SED) we adopt S/N = 5 in all bands and add to our photometric data set the monochromatic flux density of the template at $2.2$\,mm. We then fit the synthetic photometry of each galaxy by fixing the redshift first
to $z_{-}=z_{\rm true}-$$\rm\Delta$$z$, and then to  $z_{+}=z_{\rm true}+$$\rm \Delta$$z$ where $\rm \Delta$$z$ = $\epsilon$ (1+$z_{\rm true}$).  For our purposes, and based on the \zphot\ accuracy of the COSMOS field ($\epsilon = 0.005 - 0.03$ as quantified in L16) we first adopt   $\epsilon = 0.02$ and then repeat the analysis for an even more conservative case with $\epsilon = 0.05$. The comparison between the extracted $L_{\rm IR,out}$ and $M_{\rm dust,out}$ to the input values for each simulation is presented in \autoref{fig:photoz}.

Our analysis suggests that the effects $\Delta z$ has in the derivation of the FIR properties is not negligible, even for the idealised case of detailed (24 $\mu$m -- 2.2\,mm) and high quality (S/N = 5) photometric coverage. Indeed, we find that a typical value of $\epsilon = 0.02$ ($\epsilon = 0.05$),  introduces an extra scatter of $\sim$12\% (25\%) and $\sim$17\% (35\%) in the derived  $L_{\rm IR}$ and $M_{\rm dust}$, that remains rather constant with redshift (at least out to $ z = 4$). At the same time, we also find that a symmetric $\rm\Delta$z, as the one adopted in our simulations, does not inflict a noticeable systematic offset in the extracted FIR quantities ($<0.05$ dex).  

Based on these results we update the uncertainties of the inferred FIR properties of  our $z_{\rm phot}$-sample  (\texttt{ztype}=0, see \autoref{tab:catalogue}) by adding in quadrature the extra error arising from a symmetric  $\Delta z$ (assuming $\epsilon = 0.02$ for all sources) to the error budget inferred by the SED fitting procedure (photometry+model). While our correction is based on an average value of $\epsilon = 0.02$, we note that for catastrophic $z_{\rm phot}$ failures ($\epsilon \sim 0.15$) we find a systematic offset of $\leq$30\% between  the input vs output \md, while the \lir\ ratios remain uniformly scattered around unity.

\section{Completeness and Systematics} \label{sec:sims}
By construction, the photometric catalogues considered in this work combine observations that span $\sim$4 orders of magnitude in wavelength range, have  different sensitivity limits and are differently affected by source blending and confusion. The fact that we choose to draw our sample  based on the criteria described in section \autoref{sec:sample},  rather than selecting galaxies detected in a single band (i.e. a flux limited sample), necessitates a series of simulations in order to quantify possible biases, systematic effects, as well as the completeness of our sample in terms of \md\ and \lir.

\begin{figure}
\begin{center}
\includegraphics[width=0.5\textwidth]{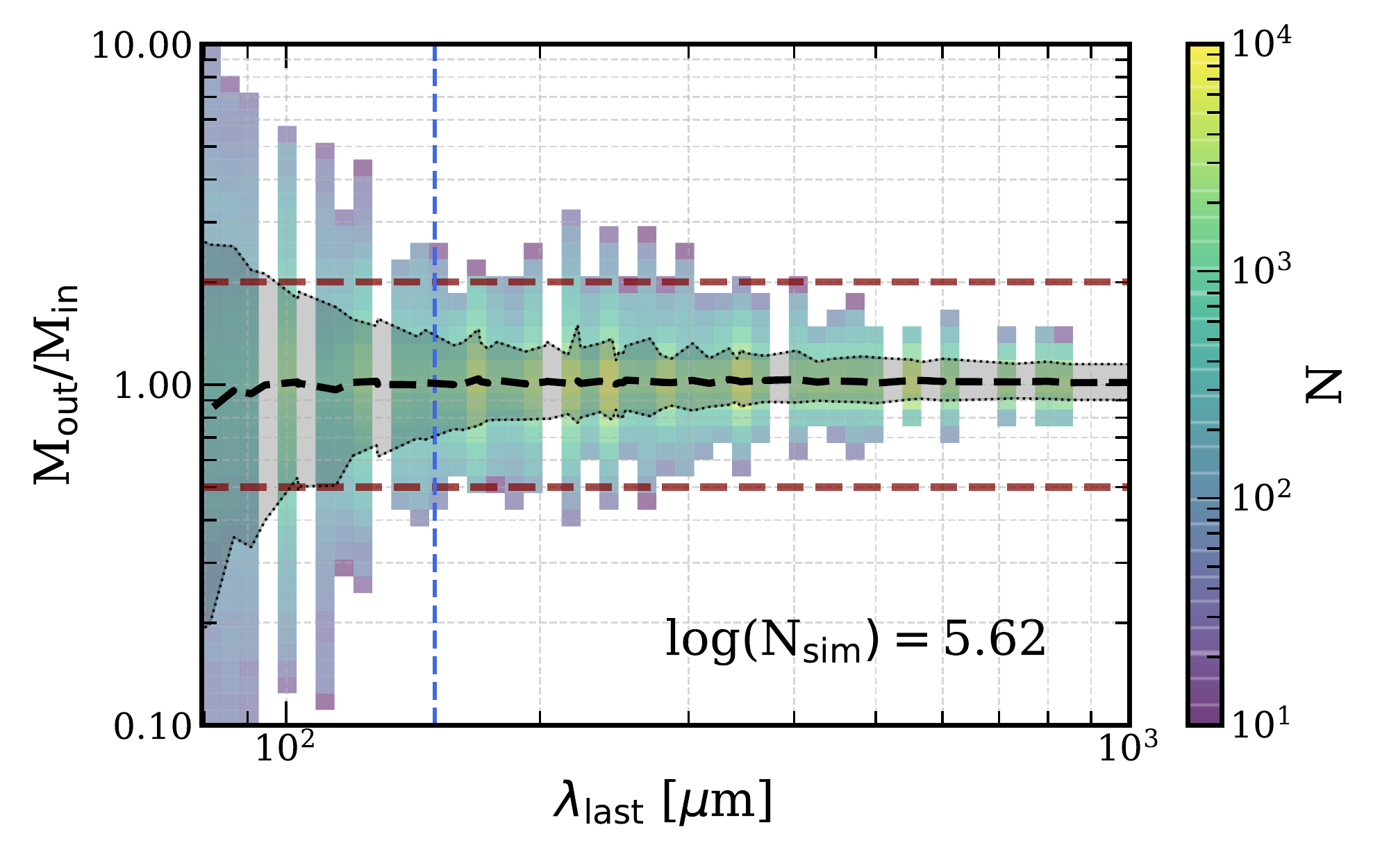}
\caption{Accuracy of the inferred \md\ estimates as a function of (rest-frame) $\lambda_{\rm last}$, parametrised by the ratio of the output to input \md\ in our simulations. The colour coding illustrates the density of the data points. The dashed black line and the shaded grey area denote the median and the 16$^{\rm th}$ and 84$^{\rm th}$ percentile confidence intervals respectively. The dashed maroon lines represent the value where the \md\ ratio is $0.5$ and $2$. The vertical blue line at 150\,$\mu$m denotes the $\lambda_{\rm last}$ onwards where for $\geq$68\% of the simulated galaxies the accuracy of the recovered \md\ is $\geq$70\%. The quantisation along the x axis is a consequence of the step-size in redshift, alongside with the available observed bands.}
\label{fig:lastdet}
\end{center}
\end{figure}

\subsection{The effect of \texorpdfstring{$\lambda_{\rm last}$}{} Cutoff} \label{sec:lambda}
It has been well established that for a robust modelling of the \md, at least one photometric data point long-wards of the FIR peak, i.e. into the R-J tail of the SED, is necessary (e.g.\ \citealt{draine07,magdis2012,berta16,scov17}). Here, we attempt to quantify how the accuracy of the derived \md\ estimates varies as a function of the rest-frame wavelength of the last available detection in conjunction with the selection criteria of our sample (i.e. at least 3 detections at $24\,\mu$m $< \lambda \leq 1200\,\mu$m). For this, we perform the following simulations.

We start by building mock IR SEDs of fixed \md\ and \lir, with 0.1 $< U_{\rm min} \leq 50$, $0 \leq \gamma \leq 0.5$ and $q_{\rm PAH}$ = 2.8\%, using the DL07 library and place them at $0.01 \leq z \leq 4.5$ with a step of ${\rm{\Delta}}z$ = 0.05. After all of the models are created, synthetic photometry is performed by convolving each mock redshifted SED with a filter transmission curve. We consider all filters redder than MIPS 24\,$\mu$m available in SDC2, for a total of $9$ bands and set all recovered fluxes to a S/N=3 significance level. At each redshift the algorithm calculates the rest-frame wavelength for each available band, producing a grid of possible rest-frame wavelengths of the last detection ($\lambda_{\rm last}$) after accounting for our selection criterion that requests the availability (and the detection) in at least two bluer bands. For each  $\lambda_{\rm last}$ it then randomly selects two additional bands bluewards of $\lambda_{\rm last}$, producing a final set of three photometric data points. We then fit each set, and 50 permutations by varying the original fluxes by 10 \%, with our code to extract \md\ and \lir\ estimates.  This procedure is then repeated for all mock SEDs and all acceptable values of $\lambda_{\rm last}$ in each redshift. 

The accuracy with which we can recover \md\ estimates for each $\lambda_{\rm last}$  is then quantified by the scatter of the output to input \md\ ratio  presented in \autoref{fig:lastdet}. As expected, we find a decreasing scatter in $M_{\rm dust,out}/M_{\rm dust,in}$ at longer $\lambda_{\rm last}$, that drops from a factor of $\sim$2 (for 68\% of the simulated galaxies) at $\lambda_{\rm last} =100\,\mu$m to a factor of $\sim$1.1 at $\lambda_{\rm last} = 400\,\mu$m. 

Based on these results, we choose to define the sub-sample of `$M_{\rm dust}$-{\it{robust}}' galaxies, for which at least one detection at $\lambda_{\rm last} \geq 150$ $\mu$m is available. This threshold was chosen as an optimal compromise between the number of the rejected sources and the precision of the derived \md\ that for $\lambda_{\rm last} \geq 150\,\mu$m is $\geq$70\%. Indeed, while $\lambda_{\rm last} \geq 150\,\mu$m is evidently not deep into the R-J, it seems that the addition of the two extra data points blueward of $\lambda_{\rm rest} = 150\,\mu$m are adequate to anchor the general shape, and eventually the \md, of the templates. 

As a sanity check for the effectiveness of our criterion, we cross match the `$M_{\rm dust}$-{\it{robust}}' sample with the A3COSMOS ALMA photometric catalogue presented in \citealt{liu19a} and refit the 233 galaxies that we find in common, adding this time the extra ALMA data point to the input photometry. The comparison of the inferred $M_{\rm dust,A3}$ to our \md\ estimates yields a very good agreement between the two estimates with a median log($M_{\rm dust,A3}$/\md) $\approx -0.04\pm0.06$,  further supporting our analysis. Nevertheless, we do identify a handful of sources for which the addition of the ALMA data point results in lower \md\ estimates by a factor of $\geq$3. An inspection of the SEDs of these extreme outliers reveals that the discrepancy originates either from possible catastrophic blending of the SPIRE\,500\,$\mu$m photometry or alternatively from over-resolved ALMA photometry \footnote{The SEDs of the most extreme outliers can be retrieved here \url{https://github.com/VasilyKokorev/sdc_ir_properties/}.\\}.

The emerging `$M_{\rm dust}$-{\it{robust}}' sample consists of 3,312 sources drawn from the same \Mstar\ and redshift distributions as  the originally 4,991 selected galaxies. Finally, we note that our simulations operate under the assumption that the DL07 models are a good representation for the FIR emission of the real galaxies. Variations in dust composition, dust emissivity and dust absorption coefficients that could result in systematic offsets in the inferred \md\ \citep[e.g.][]{magdis2012,dale12,berta16,scov17} cannot be modelled with our approach. As is the case for any other \md\ analysis - the relative rather than the face value estimates bare more physical significance.

\subsection{Limiting \texorpdfstring{$M_{\rm dust}$}{} and \texorpdfstring{$L_{\rm IR}$}{}} \label{sec:limmd}
We now  attempt to compute the completeness threshold of our sample in terms of $M_{\rm dust}$ and \lir\ as a function of redshift. Again, we build a grid of mock IR SED in the $z=0-5$ range using the same approach as described above. However, instead of considering the full range of possible \avu $\propto$ \lir/\md\ values of the DL07 models, the constructed templates this time follow the \avu\ -- $z$ relation of MS galaxies presented in \citet{bethermin15}. At each redshift we then create a grid of templates normalised to log($M_{\rm dust}$/\msol) = $6-10$ in steps of 0.1. \par
As before, the templates are used to  produce synthetic photometry for each template in all bands available in the SDC2 catalogue. For each band we then adopt an rms based on the depth of the corresponding survey at each wavelength in the COSMOS field (\citealt{sdc2}), and impose the same selection  criteria to the mock photometry as those applied to the real catalogues. Following \autoref{sec:lambda} we also request that the simulated sources have  $\lambda_{\rm last} \geq 150\,\mu$m. If a galaxy of given \md\ fulfils these criteria at a given redshift, the algorithm moves to a lower \md\ until the object is rejected by our selection. By following these steps at different redshifts we thus obtain a limiting \md\ as a function of $z$, that we coin  lim($M_{\rm dust}$)($z$). In the process we also consider the scatter of the \avu\ -- $z$ relation of \citealt{bethermin15}, in order to account for the variation of \avu\ among MS galaxies at a given redshift. The derived lim($M_{\rm dust}$)($z$) can then be converted to  lim($L_{\rm IR}$)($z$), via:
\begin{equation}
\langle U \rangle = \frac{L_{\rm IR,DL07}}{125\, M_{\rm dust}},
\end{equation}
as described in \citet{dl07}.

 We also repeat our simulation for SBs, by fixing   $\langle U \rangle = 40$ (e.g.\ \citealt{magdis2012,tan14,bethermin15}). We note however that the \avu\ of SBs can vary substantially to lower or higher values (e.g.\ \citealt{magdis2012,magdis17,schreiber18,jin19,cortzen20}). Therefore, the chosen \avu\ = 40 is only representative, but not necessarily unique. Nevertheless, \avu\ $<40$ templates are represented in the simulation of the MS galaxies while galaxies with \avu\ $>40$ are rather rare.

\begin{deluxetable*}{cccccccccc}
\tablecaption{\label{tab:general-fir} Properties of the galaxy sample selected in \autoref{sec:sample}. With the exception of redshift, the other quantities are derived via SED fitting. Values are presented as the median and a double sided $68\%$ confidence interval.\\}
\tablehead{%
	& \# &  $z$ 	&  $L_{\rm IR,total}$ $^{a}$ 	& $L_{\rm IR,DL07}$ $^{b}$ & SFR$^{c}$ & $M_{\rm dust}$ & $M_\ast$ $^{d}$ & $f_{\rm AGN}$ & \avu \\
& Galaxies &    & [$10^{12}\,\rm{L}_{\odot}$]  & [$10^{12}\,\rm{L}_{\odot}$] &  [M$_{\odot}$ yr$^{-1}$] &  [$10^{8}\,\rm{M}_{\odot}$] & [$10^{10}\,\rm{M}_{\odot}$] &  [\%] &}
\startdata
\\
COSMOS & 4,406 &	0.88$^{+1.09}_{-0.57}$ & 0.45$^{+2.17}_{-0.40}$ & 0.44$^{+2.10}_{-0.39}$ & 43.96$^{+209.93}_{-38.88}$ & 2.73$^{+16.66}_{-2.08}$ &4.17$^{+5.83}_{-2.89}$ & 0.86$^{+2.44}_{-0.60}$ & 10.12$^{+29.55}_{-7.72}$ \\
\\
GOODS-N & 585 &	1.01$^{+1.02}_{-0.53}$ & 0.35$^{+1.60}_{-0.28}$ & 0.34$^{+1.53}_{-0.27}$ & 33.84$^{+152.74}_{-27.02}$ & 2.60$^{+10.34}_{-1.76}$ &3.56$^{+6.03}_{-2.32}$ & 2.60$^{+4.74}_{-2.14}$ & 9.38$^{+29.82}_{-6.95}$ \\
\\
All & 4,991 & 0.90$^{+1.08}_{-0.58}$ & 0.44$^{+2.09}_{-0.38}$ & 0.42$^{+2.04}_{-0.37}$ & 42.30$^{+203.88}_{-37.06}$ & 2.71$^{+15.71}_{-2.04}$ &4.07$^{+5.93}_{-2.72}$ & 0.93$^{+3.06}_{-0.67}$ & 10.00$^{+29.68}_{-7.58}$ \\
\\
\enddata
\begin{tablenotes}
\item[a] \footnotesize{Computed over a linear combination of AGN+DL07 best fit templates.}
\item[b] \footnotesize{Only considering the best fit DL07 template.}
\item[c] \footnotesize{Computed from $L_{\rm IR,DL07}$.}
\item[d] \footnotesize{Taken from the parent catalogue.}
\end{tablenotes}
\end{deluxetable*}

\begin{figure}
\begin{center}
\includegraphics[width=.5\textwidth]{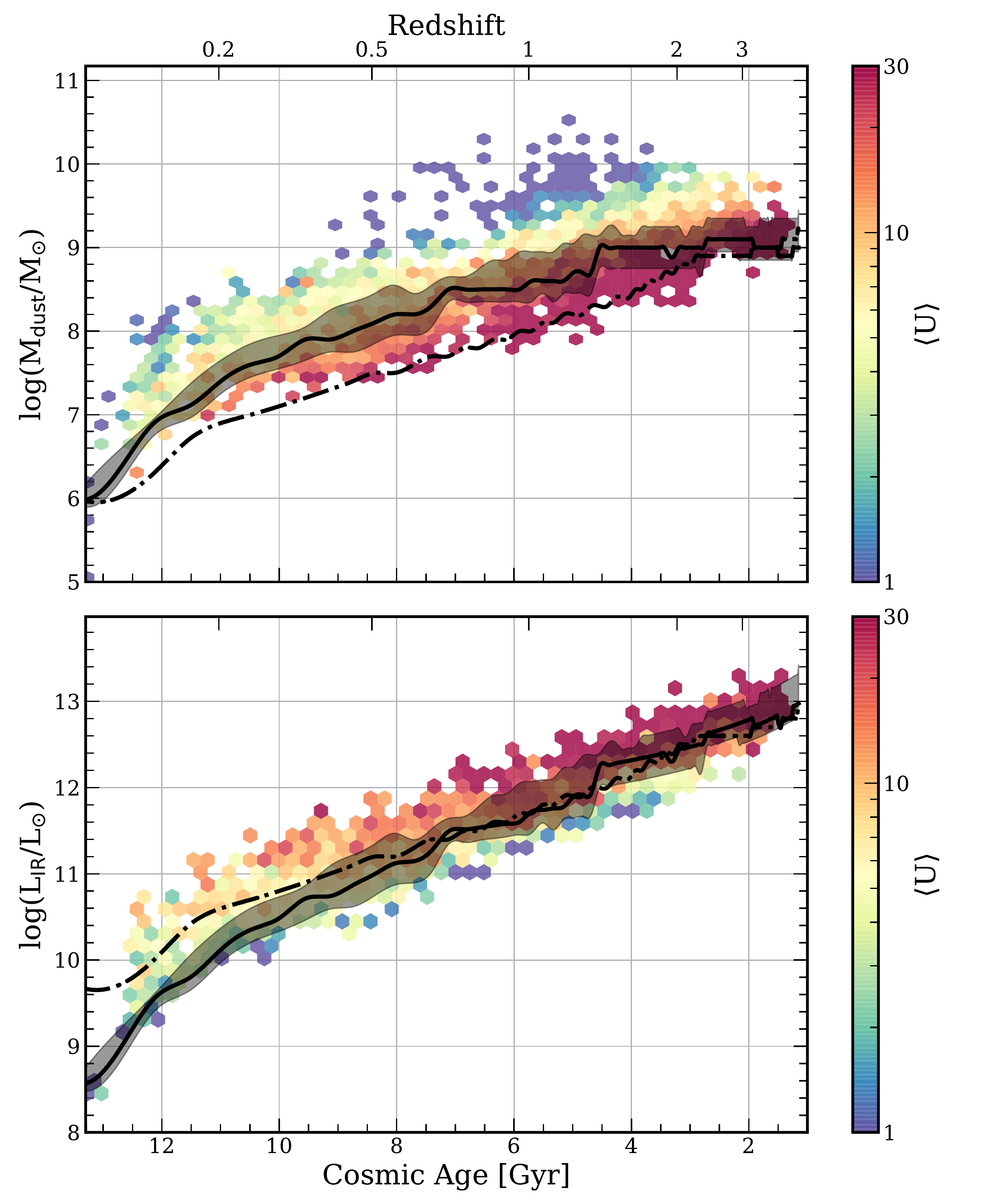}
\caption{Simulated evolution of the lim($M_{\rm dust}$)($z$) (top) and lim($L_{\rm IR}$)($z$) (bottom), as described in \autoref{sec:limmd}. The black line represents the derived trend for MS galaxies and the dashed-dotted line shows the same relations for SBs (\avu\ $\sim40$). The shaded regions define the $16^{\rm th}$ and $84^{\rm th}$ percentile confidence intervals, based on the scatter of the \avu\ -- $z$ relation from \citet{bethermin15}. The hexagonal bins contain the inferred parameters of the `\md-{\it{robust}'} sample (i.e. at least one detection at $\lambda \geq 150\,\mu$m), colour coded by mean radiation field intensity \avu.}
\label{fig:lim_md}
\end{center}
\end{figure}

\begin{figure*}
\begin{center}
\includegraphics[width=1\textwidth]{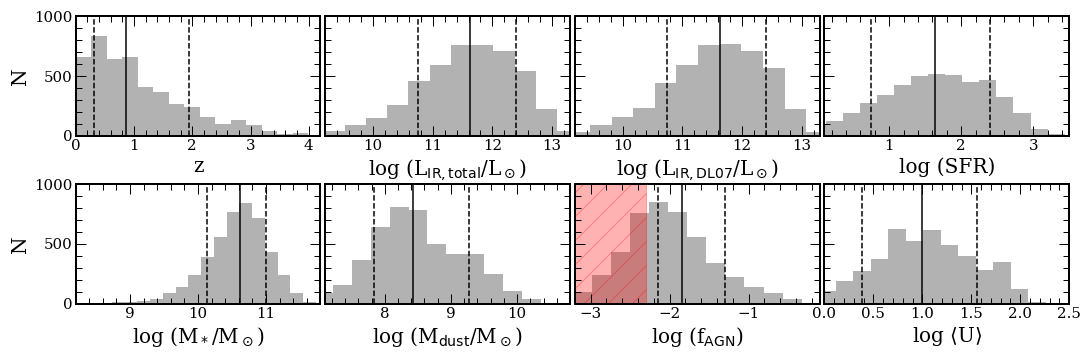}
\caption{Distribution of the inferred IR properties of the COSMOS and GOODS-N samples. 
 With the exception of $z$ and $M_*$, these properties are the output of our SED fitting code (see \autoref{sec:sed}).  The solid black and dashed lines represent the median and the 68\% confidence interval respectively. The hatched red region on the $f_{\rm AGN}$ histogram highlights the range where estimates are not reliable (i.e., $f_{\rm AGN}<0.005$). }
\label{fig:hist_sample}
\end{center}
\end{figure*}

The results of our simulations are presented in \autoref{fig:lim_md}, where we show the derived \md\ and \lir\ as a function of $z$ for the whole sample, along  with the evolution of lim($M_{\rm dust}$) and lim($L_{\rm IR}$). We see that the limiting \md\ increases towards high-$z$, peaking at $z\sim2$ and remains flat afterwards, signifying  that the balance between cosmological dimming and negative K - correction is achieved beyond that point. Following the black line we could infer the \md\ threshold above which our sample should be 100 \% complete, assuming an MS-like population of galaxies. However, since our sample is not limited to MS galaxies we naturally also find sources that fall below our limiting \md\ track. As we will discuss later, these are predominately starbursting galaxies that exhibit elevated $\langle U \rangle$  with respect to the MS. Our secondary lim($M_{\rm dust}$) trend for an SB-like population displays that with the SDC2 detection limits it is possible to detect a low $M_{\rm dust}$ object only if it is also a starburst. \par
We also find a similar trend for MS galaxies when considering the 
evolution of lim($L_{\rm IR}$). In this case, however we do not observe a flattening at $z\sim2$ and the trend continues to rise into the early Universe. The balance between cosmological dimming and the negative K - correction is not being achieved here, since the wavelengths that are required to reliably constrain the $L_{\rm IR}$ are positioned to the left of the FIR peak.\par
Admittedly, depths of FIR surveys are not the only limiting factors of sample selection. A requirement to have a photometric redshift and $M_*$, would mean that the optical photometry has to be sufficiently sampled, to allow such an analysis. Moreover, the deblending procedure itself goes through various selection stages, including both brightness and mass cuts. Various IR studies \citep{wang16,franco18} have revealed substantial populations of optically dark sources at $2<z<4$, that are otherwise bright in IRAC and FIR bands, which would be unintentionally excluded from our analysis. Moreover, even at low-$z$, objects that are faint in the $K$-band would also be missed. Indeed, a combination of these factors creates significant obstacles in our completeness analysis, we address this in more detail in \autoref{sec:dmf}.

\begin{figure}
\begin{center}
\includegraphics[width=0.49\textwidth]{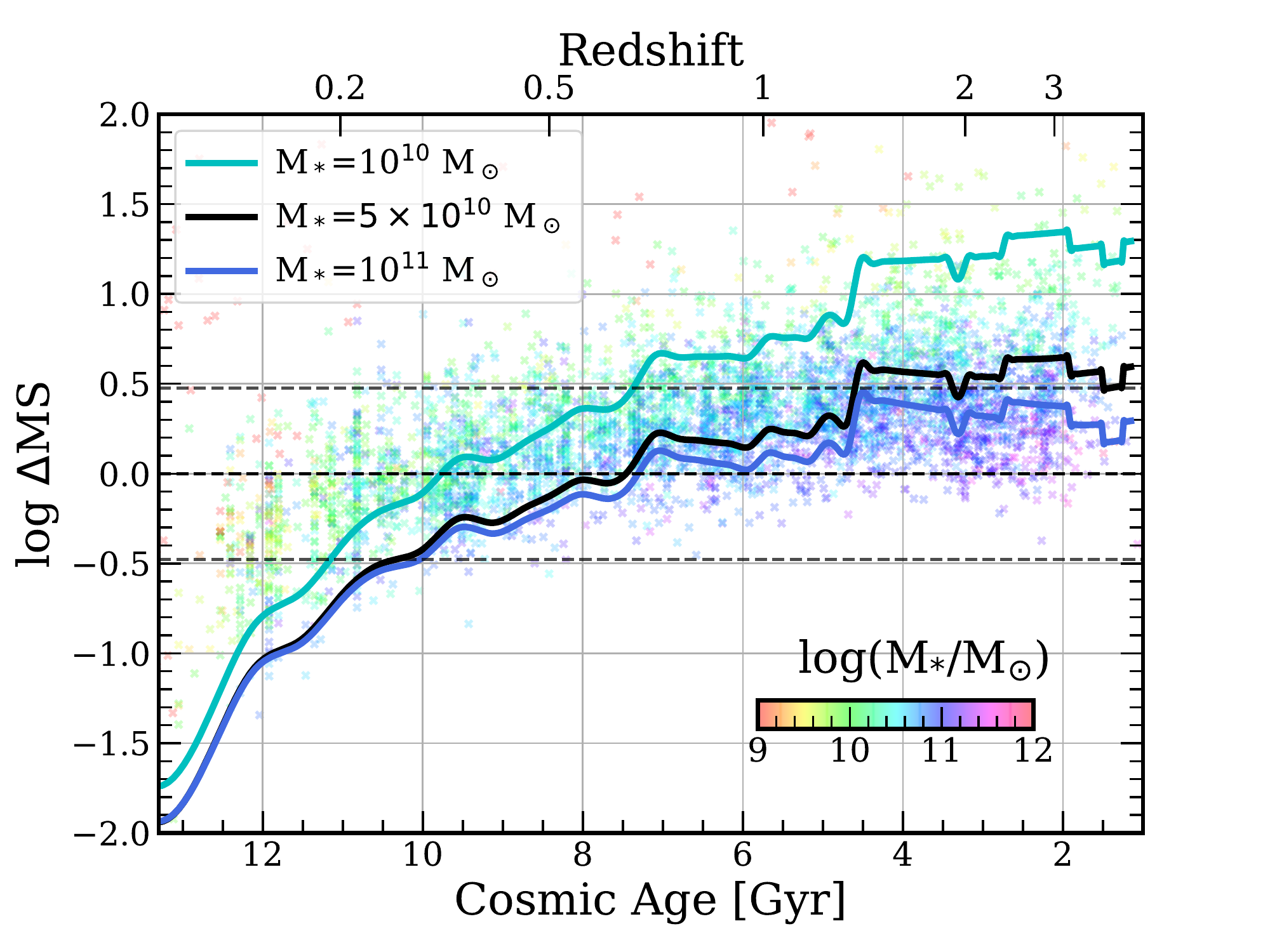}
\caption{Position of our sources with respect to the MS as a function of cosmic age and redshift. Points are colour coded according to the $M_*$.  The dashed black and gray lines denote the MS and its 0.5$\,$dex scatter. The solid coloured lines correspond to the $\Delta$MS detection limit as computed based on the inferred $lim(L_{\rm IR})$ (see \autoref{sec:limmd}), and assuming \Mstar= 10$^{10}$, 5$\times10^{10}$ and $10^{11}$ \msol.}
\label{fig:delta_ms}  
\end{center}
\end{figure}

\begin{figure*}
\begin{center}
\includegraphics[width=1.\textwidth]{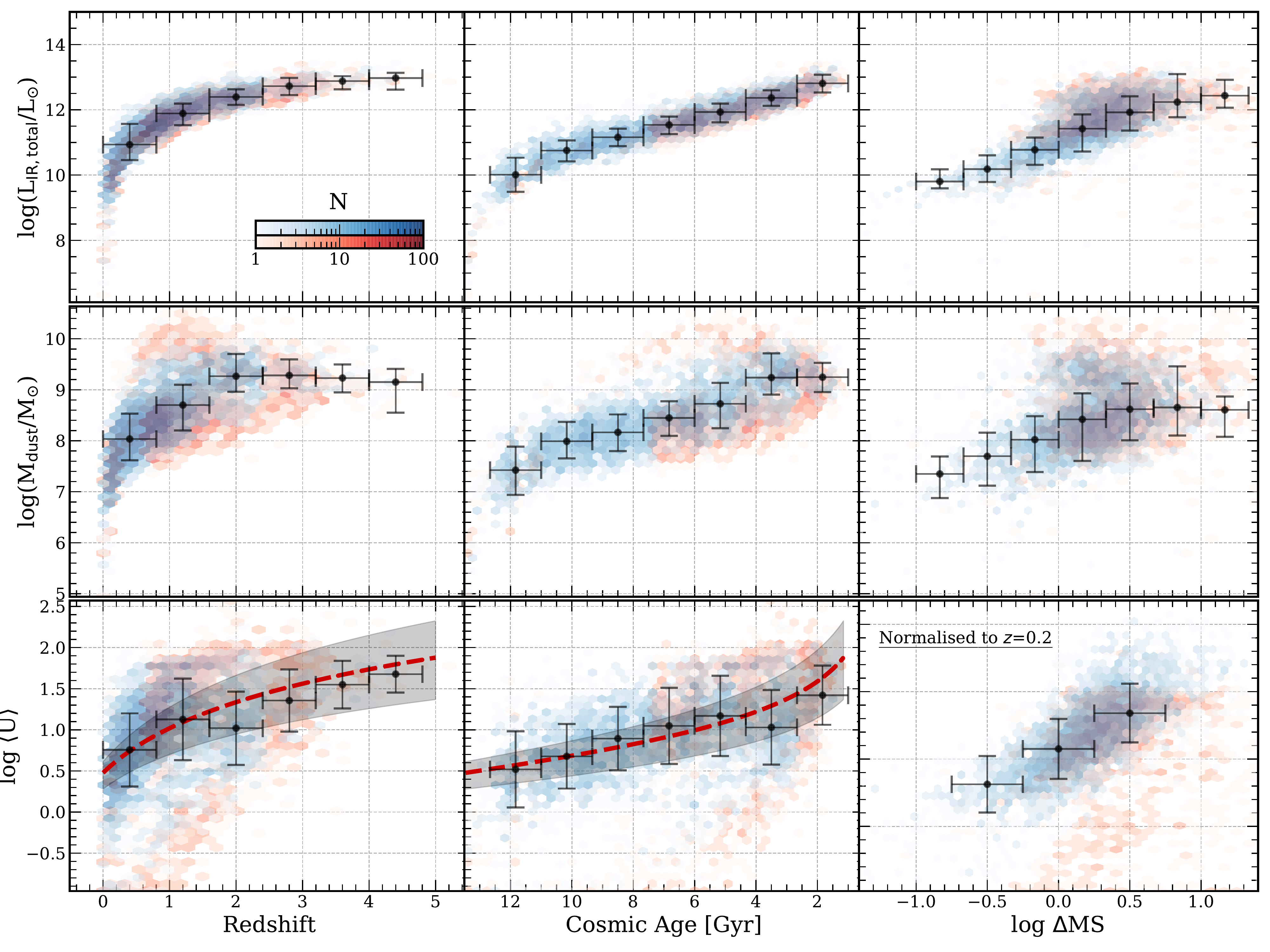}
\caption{Evolution of general FIR properties, as computed with \texttt{Stardust}, as a function of $z$, cosmic age and $\Delta$MS. The hexagonal bins are normalised by the number count, and contain the `\md-{\it{robust}}' sample in blue, and the objects that were removed after the quality cut in red. For the `\md-{\it{robust}}' we show the binned median points, with their $y$-uncertainty corresponding to the 16$^{\rm th}$ and 84$^{\rm th}$ percentile intervals and $x$-uncertainty to the bin width. The dashed red lines and shaded regions correspond to the \avu\ -- $z$ relation for MS galaxies from \citet{bethermin15}. }
\label{fig:test}
\end{center}
\end{figure*}

\section{Far-infrared Properties of GOODS-N and COSMOS galaxies} \label{sec:genprop} 
Using our newly developed code presented in \autoref{sec:sed}, we extract the FIR and UV-optical properties for all 4,991 galaxies from the SDC1 and SDC2 that meet our selection criteria as listed in \autoref{sec:sample}. Moreover, since both input catalogues contain $M_*$ estimates provided by either \texttt{LePhare}, \texttt{EAZY} or \texttt{FAST}, we are able to carry out a comparison of these $M_*$ to the ones derived by \texttt{Stardust}. We find that the stellar masses are consistent with one another and direct the reader to \autoref{sec:appendix-c} for a more detailed comparison between the two methods, as well as to \texttt{EAZY} derived $M_*$. Despite the similarities between the available and derived $M_*$, in our subsequent analysis we will utilise the $M_*$ from the parent catalogue, unless it is specified otherwise. This is done to preserve the original mass cuts described in \citet{sdc1} and \citet{sdc2}, and therefore the mass completeness and homogeneity of the catalogue.

In total, out of 4,991 sources, we find 21 that we consider to be catastrophic fits ($\chi_\nu^2>100$), these only comprise 0.4 \% of the entire output catalogue and are subsequently removed. The average $\chi^2$ per degree of freedom of the entire dataset was computed to be equal to 0.98. The distribution of the FIR properties of the whole sample, their medians and associated uncertainties are presented in \autoref{fig:hist_sample} and summarised in \autoref{tab:general-fir}. The catalogue containing the extracted FIR properties is described in the Appendix and is publicly available along with the best fit SED for each object.

We also calculate the position of the galaxies in our sample with respect to the MS, by converting the AGN-free $L_{\rm IR,DL07}$ estimates to SFR \citep{Kennicutt98} and using the functional form of the main-sequence as presented in \citet{schreiber15}, accounting for the fact that they use a \citet{salpeter} IMF. The distribution of  $\Delta$MS= SFR/SFR$_{\rm MS}$ as a function of redshift and stellar mass is presented in \autoref{fig:delta_ms}. We define the boundary between the star-forming and quiescent galaxies at log$\Delta$MS=$-$0.5 dex and between MS and SBs at log$\Delta$MS=0.5 dex, which in linear space corresponds to $\times$3 below/above the MS. Quite naturally, for decreasing \Mstar\ and increasing redshift, our sample is progressively restricted to galaxies that lie above the MS. This is shown by the tracks in \autoref{fig:delta_ms} that indicate the limiting $\Delta$MS for fixed \Mstar\ as a function of redshift that is reached by our data, after  converting the inferred $lim$(\lir) to $lim$(SFR). Nevertheless, we find that the majority of our sources are classified as MS galaxies ($69\%$), with the remaining objects either considered to be undergoing a phase of `bursty' star-formation ($26\%$) or being passive galaxies ($5\%$). 

\par
As a final sanity check we additionally fit the same sources with \texttt{CIGALE}, by utilising DL07 models and similar sets of optical and AGN templates. We find that the output parameters as derived from the  two codes are in good agreement and defer the reader to \autoref{sec:appendix-d} for a more detailed comparison. 

\subsection{The `\texorpdfstring{\md-{\it{robust}}}{}' Sample}

We now focus on the FIR properties of the `\md-{\it{robust}}' galaxies described in \autoref{sec:lambda}, that should represent the most reliable sample for the exploration of the dependency of the \lir, \md, and \avu\  on redshift, cosmic age, and $\Delta$MS. The emerging results are presented in \autoref{fig:test}, where for completeness and to facilitate comparisons, we also include the inferred properties of the full sample. \par
Both \lir\ and \md\ are found to increase smoothly as a function of $\Delta$MS. At the same time, we also find that for MS galaxies \avu\ evolves  as  $(3.2\pm1.3)\times(1+z)^{1.2\pm0.3}$, which is in excellent agreement with the stacking analysis of \citet{bethermin15}. The fact that our individually detected galaxies appear to follow the same  \avu\ - $z$ relation as the much deeper stacked ensembles, reinforces the notion that the adopted `\md-{\it{robust}}' sub-sample does not introduce a significant bias towards colder objects.\par
Since \avu\ is proportional to $L_{\rm IR,DL07}$/$M_{\rm dust}$, and also a proxy for $T_{\rm d}$,  our analysis provides further  evidence that dust in MS galaxies becomes warmer towards higher redshifts, a trend that has already been recovered in previous studies  \citep[most of them based on stacking analysis, see e.g.][]{magnelli2014,schreiber15,davidzon18}. 
Similarly, our data also confirm a progressive increase of \avu\ (or $T_{\rm dust}$) with an increasing elevation above the MS (e.g. \citealt{magdis17,jin19}).

It is worth noticing that the full sample follows the same general trends albeit with a considerably larger scatter ($\sim \times2$) in \md\ and \avu. The reduced scatter for the `\md-{\it{robust}}' sub-sample is driven by the imposed $\lambda_{\rm last}\geq150\,\mu$m selection criteria that primarily removes the locus of sources with very cold fitting solutions (\avu\ $\lesssim 1$). We highlight that the rejection of these objects should not introduce a bias in our sample since such low \avu\ values are more indicative of poor photometric coverage/quality (lack of available data point in the R-J) rather than of realistic, extremely cold ISM conditions. However, we note that not all the extremely cold solutions have been removed from  the `\md-{\it{robust}}' sample by our selection, as $\sim$200 objects with \avu\ $<1$ meet our $\lambda_{\rm last}>150\,\mu$m criterion. These can be easily identified in the \avu\ $-$ $z$ plot and as the outliers populating the secondary blue cloud of points in the $M_{\rm dust}-\Delta$MS plot in \autoref{fig:test}. As we will discuss later, these could be sources with unreliable $z_\mathrm{phot}$ estimates, failures of the deblending in the SPIRE bands, or, more interestingly, gas giants or very compact galaxies with optically thick FIR emission. 

\subsection{Cold vs Warm Dust}
The SED decomposition introduced in \autoref{sec:sed} can also provide constrains on the relative contribution of the warm (PDR, $L^{\rm warm}_{\rm IR}$, $M^{\rm warm}_{\rm dust}$) and cold (diffuse, $L^{\rm cold}_{\rm IR}$, $M^{\rm cold}_{\rm dust}$) ISM components to the total $L_{\rm IR}$ output and the total $M_{\rm dust}$ budget of the galaxies in our sample. In particular, it is worth investigating if and how the relative contribution of the components varies as a function of $\Delta$MS. Indeed, if SBs are experiencing elevated star-formation activity per surface area \citep{elbaz2011,elbaz18,valentino20} then one would expect to see an increased fraction of \lir\ (and $M_{\rm dust}$) originating (and being heated) from the `PDR' component, where the radiation intensity ranges from $U_{\rm min}$ to $U_{\rm max}$ \citep{dl07}. \par

\begin{figure*}
\begin{center}
\includegraphics[width=1\textwidth]{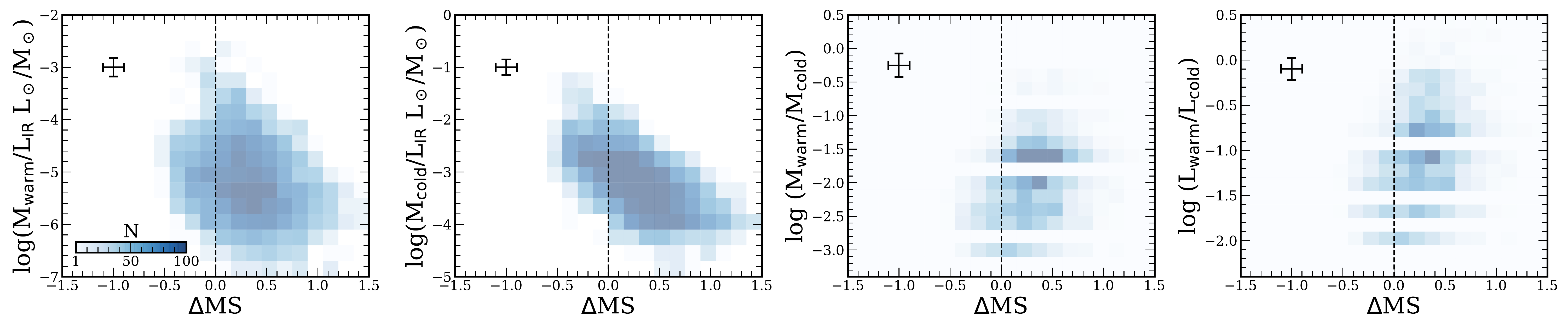}
\caption{Comparison of the properties of the warm (PDR) and cold (diffuse) dust components of the ISM as a function of $\Delta$MS. From left to right the  panels show the warm and cold dust mass components weighted by the total \lir, the fraction of warm to cold $M_{\rm dust}$ and the fraction of `PDR' to diffuse ISM IR output. The bins are coloured based on number density of the data points. We show a typical uncertainty on the plotted parameters in the top left corner of each panel.}
\label{fig:cold_warm}
\end{center}
\end{figure*}

In \autoref{fig:cold_warm} we plot the inferred properties of the warm and cold ISM components as a function of $\rm \Delta$MS. We find that for a fixed $L_{\rm IR}$ (or equally SFR), SBs tend to have lower amounts of $M^{\rm cold}_{\rm dust}$, with $M^{\rm cold}_{\rm dust}$/$L_{\rm IR}$ showing a tight anti-correlation with $\rm \Delta$MS. The same, however, does not apply to $M^{\rm warm}_{\rm dust}$/\lir, which exhibits a significantly larger scatter and only a very weak dependence on $\rm \Delta$MS. This is a consequence of an increasing fraction of warm to cold \md\ and \lir\ between MS and SBs galaxies (\autoref{fig:cold_warm}). 

The observed trends suggest that compared to MS galaxies, SBs have a larger fraction of their total \md\ exposed to the intense stellar radiation fields of the PDRs, in agreement with expectations discussed above. Our result could indeed reflect an increase in the compactness of the star-formation activity for increasing distance from the MS as suggested in recent high resolution studies. Finally, under the assumption that $M^{\rm cold}_{\rm dust}$ is proportional to \mgas, and \lir\ is proportional to SFR, our results point to enhanced star-formation efficiencies and shorter gas depletion time scales for sources residing above the MS, as already reported in the literature (e.g.\ \citealt{tacconi10,tacconi20,daddi10,magdis2012,magdis17,sargent14,silverman18}).

\section{Dust to Stellar Mass Relation and Dust Mass Functions} \label{sec:dms}
As discussed in \autoref{sec:intro}, constrains on the evolution of $f_{\rm dust}$ and the DMF are key towards a better understanding of  dust production and destruction mechanisms at different epochs. Within this context, we explore how the current data set traces the evolution of $f_{\rm dust}$ and use it to  characterise the DMFs at various redshifts.

\subsection{The Evolution of the Dust Mass Fraction}
To infer the evolution of $f_{\rm dust}$ we adopted the formula described in \citet{liu19b}, which parametrises \fdust\ in terms of $z$, $M_*$ and $\Delta$MS.  Compared to more simple log-space linear fitting models (e.g.\ \citealt{scov17}), this formulation recovers trends that are more physically meaningful and also explores how these parameters are covariant and degenerate with each other in a multi-dimensional fitting space. As an initial check, we performed a Spearman correlation test and found $M_{\rm dust}$ to be mildly correlated with log$\Delta$MS ($\rho=0.40$) and strongly correlated with $M_{*}$ and $t_{\rm age}$ ($\rho=0.63$ and $-0.80$ respectively). We consider the following functional form:

\begin{align}
\mathrm{log}( f_{\rm dust})= &(a_0+a_1\, \mathrm{log}(M_*/10^{10} M_\odot))\, \mathrm{log}\Delta \mathrm{MS} \nonumber\\ 
&+b\, \mathrm{log}(M_*/10^{10}M_\odot))\label{eq:1} \\
&+(c_0+c_1\, \mathrm{log}(M_*/10^{10}M_\odot))\, t_{\rm age} (z) \nonumber \\
&+d \nonumber
\end{align}

\noindent where $t_{\rm age}$ is the cosmic age at a given redshift in Gyr and $M_*$ is the stellar mass in M$_\odot$. For fitting we use the \texttt{Python} package - \texttt{scipy.optimize.curve$\_$fit}, which finds the solution based on the least-squares method. We also consider how the extreme outliers can affect our results and thus only fit the medians in a given redshift bin. The best fit values are as follows:
\begin{align*}
&a_0=-0.90 &\quad a_1=1.31  \\
&b= -0.98  &\quad c_0=-0.23  \\
&c_1= 0.11  &\quad d=-0.64,
\end{align*}

\noindent with the uncertainties being computed from the covariance matrix. We then used the functional form given by \autoref{eq:1} to re-normalise all galaxies to lie on the main sequence ($\Delta$MS = 1), and $M_*=5\times10^{10}$ M$_\odot$, in order to directly compare with our best-fit function in 2 dimensions.\par
\begin{figure}
\begin{center}
\includegraphics[width=0.45\textwidth]{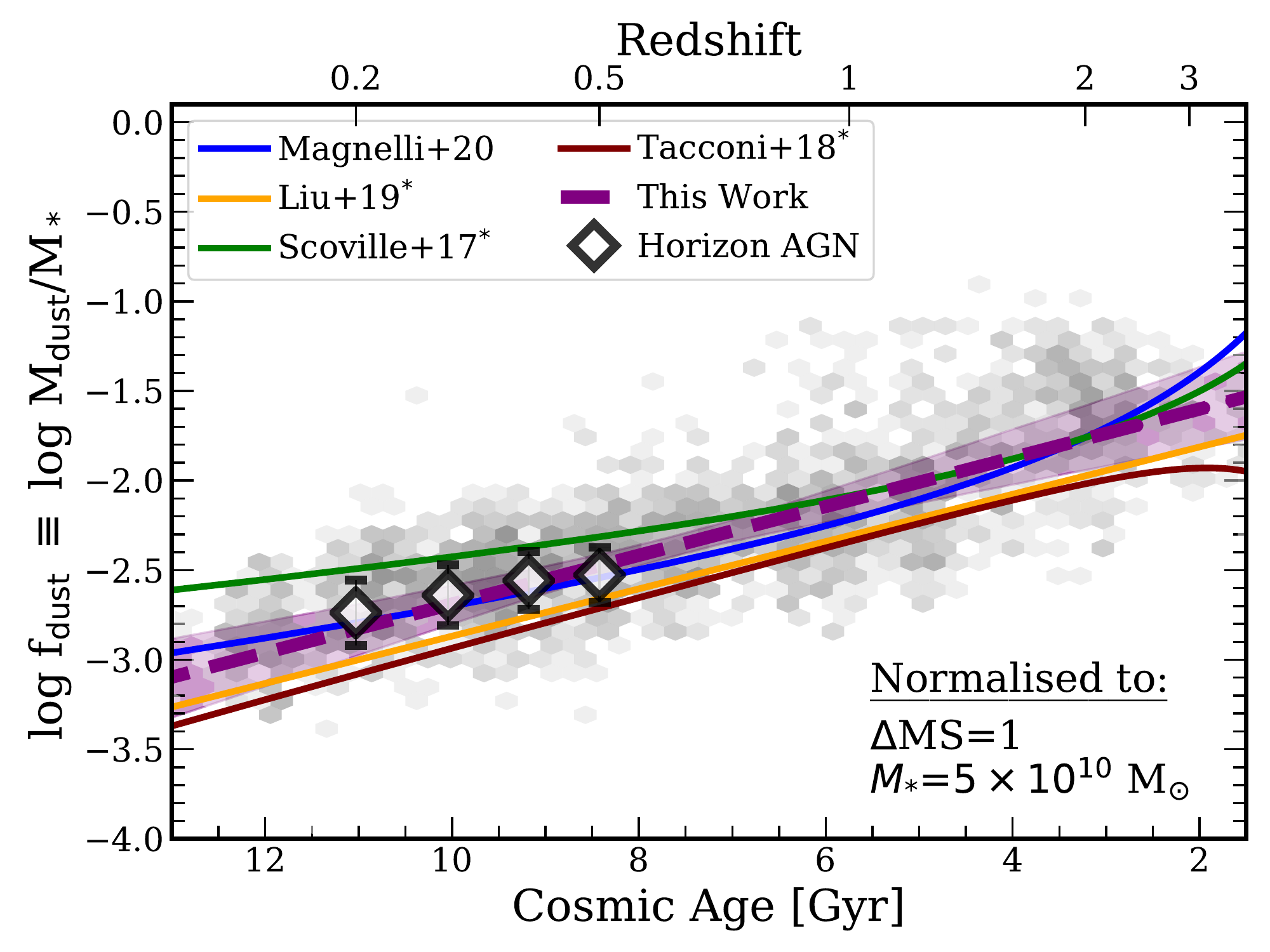}
\caption{Derived relations for $f_{\rm dust}$ as a function of $z/t_{\rm age}$. The dashed purple line shows the fit to our data, while solid coloured lines display literature results. The shaded purple region denotes the 16$^{\rm th}$ and 84$^{\rm th}$ percentile confidence intervals of our fit. Starred labels denote literature calculations where a direct comparison was not available and a $\delta_\mathrm{GDR}=100$ was assumed. The grey hexbins contain the data from `\md-{\it{robust}}' sample, and are normalised by the number count. Both the data and the derived relations have been re-scaled to $\Delta$MS=1 and $M_*=5\times10^{10}$ M$_\odot$.
White diamonds show median positions of the Horizon AGN star-forming galaxies at that redshift, normalised in the same way as our data.}
\label{fig:trends-dust}
\end{center}
\end{figure}

The normalised data and the best fit relation presented in \autoref{fig:trends-dust} are in very good agreement with the collection of similar trends drawn from the literature 
\citep[e.g.][]{scov17,tacconi18,liu19b,magnelli20}. We note that any apparent discrepancies between the slope and the normalisation of our recovered relation to that of \citet{scov17} and \citet{magnelli20} are driven by the model parametrisation, as for the latter the multi-variable functional forms do not consider the covariance between the fitted values.

\begin{figure*}
\begin{center}
\includegraphics[width=0.85\textwidth]{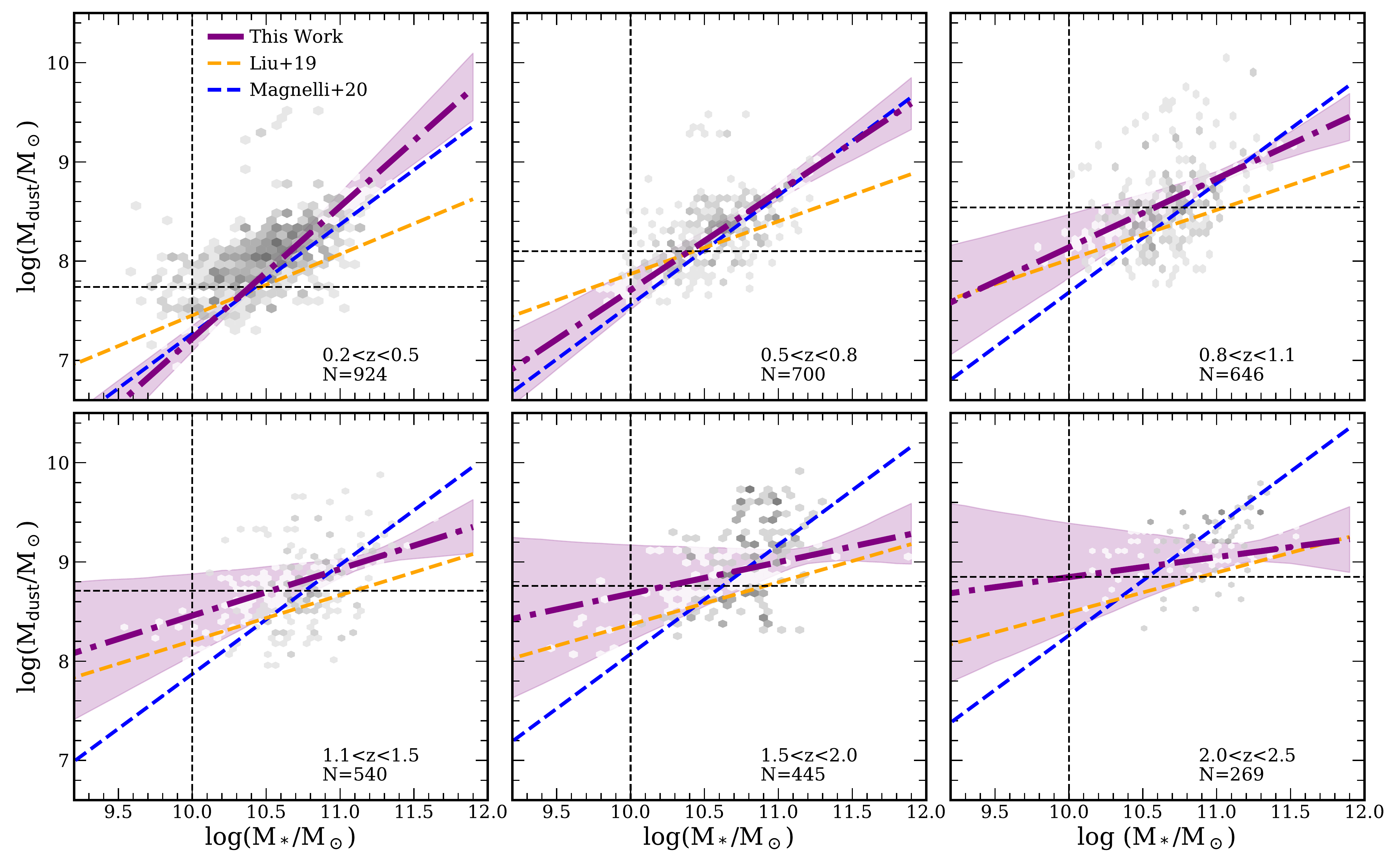}
\caption{$M_{\rm dust}$ as a function of $M_*$ in six redshift bins. The dashed-dotted purple line represents our best fit from \autoref{eq:1} that was collapsed to a single dimension, with $\Delta$MS=1 and $z=\langle z_{\rm bin} \rangle$. The shaded purple regions denote the 16$^{\rm th}$ and 84$^{\rm th}$ percentile confidence intervals of our fit. The orange and the blue lines show the relations derived in \citet{liu19b} and \citet{magnelli20}, respectively. The dashed black lines represent the detection limits of the original catalogue, in $M_*$ (vertical), and the lim($M_{\rm dust}$) that we compute in \autoref{sec:limmd} (horizontal).}
\label{fig:dmf1}
\end{center}
\end{figure*}

In \autoref{fig:dmf1} we also show how the one dimensional relation between \md\ and $M_*$ compares to the multi-dimensional fit within six redshift bins. Since our analysis can be affected by the completeness of our sample in terms of \Mstar\ and \md, we also consider the underlying selection effects based on our lim($M_{\rm dust})(z)$ derivation and assuming that our catalogue is complete at \Mstar $> 10^{10}$\,\msol. We find that for a fixed redshift range, the inferred \md $-$ \Mstar\ relation shadows the multi-parameter fit up to $z\sim 1.1$, in moderate agreement with the trends reported in \citet{liu19b} and \citet{magnelli20}. However, the increasing incompleteness and the low number statistics do not allow us to extend our analysis at higher redshifts.

Based on these results we will limit the subsequent DMF analysis to the $0.2<z<1.1$ range.

\subsection{Dust Mass Functions}
\label{sec:dmf}
After obtaining the functional form for $f_{\rm dust}(t_{\rm age},M_*,\Delta\rm{MS})$, we are now in position to examine the shape and the evolution of the DMF with redshift.
For this we will restrict our analysis to the COSMOS sample (SDC2), as we do not have enough statistics or coverage of the GOODS-N field to reliably constrain the DMF.

\par The 'Super-Deblending' procedure that went into producing the SDC2 catalogue creates a significant obstacle when attempting to consider all of the incompleteness effects of the sample. The objects that end up in our sample  go through several selection stages, both before and after the deblending procedure. These include both the brightness and mass cuts of the parent catalogue \citep{laigle2016}, the availability of infrared coverage \citep{sdc2} and the selection criteria imposed in our study. As such, it is not possible to robustly assess the properties and the number of objects that end up being `lost'. Therefore, we select an alternative approach in computing the DMF for our objects, namely utilising the derived \fdust\ parametrisation along with the available SMFs in the literature. Later in this section, we also attempt to account for the incompleteness effects and compare the SMF-derived DMF to the observed number density of galaxies per $M_{\rm dust}$ bin.

\subsubsection{DMF from SMF}
For our analysis we adopt the SMF computed by \citet{davidzon2017}, which covers the entire COSMOS field and the $z=0.2-4$ range. Their mass function already accounts for the Eddington bias, so we do not need to consider any additional corrections.  Since the vast majority of galaxies in our sample are star-forming, we adopt the derived parameters for the `active' SMF only. \par
The galaxy mass function is normally expressed as a Schechter function \citep{schechter}, which in logarithmic form can be written as:
\begin{equation}
\begin{split}
\Phi (\mathrm{log}M) d(\mathrm {log}M)=\mathrm{ln}(10) \, \mathrm{exp}(-M/M^*)\\ \times \Phi^*(-M/M^*)^{(\alpha+1)} d(\mathrm {log}M_*)
\end{split}
\end{equation}
where $\alpha$ is the slope of the faint-end, $\Phi^*$ is the normalisation and $M^*$ is the characteristic mass, indicating the position of the `knee' of the Schechter function. 
To convert the SMF to DMF, we first postulate that the number of galaxies in a given redshift bin is the same, regardless of whether we integrate over $M_*$ or $M_{\rm dust}$, namely:
\begin{equation*}
\int_0^\infty \Phi (\mathrm {log}M_*) d(\mathrm {log}M_*)= \int_0^\infty \Phi (\mathrm {log}M_{\rm dust}) d(\mathrm {log}M_{\rm dust}), 
\end{equation*} 
the integrands can be re-arranged to obtain:
\begin{equation*}
\Phi (\mathrm {log}M_{\rm dust})= \Phi (\mathrm {log}M_*) d(\mathrm {log}M_*)/d(\mathrm {log}M_{dust}).
\end{equation*}
We then differentiate \autoref{eq:1}, to obtain $d(\mathrm {log}M_*)/d(\mathrm {log}M_{\rm dust})=(b+c_1\times t_{\rm age}+1)^{-1}$. In order to perform the final conversion we also transform all the $M_*$ bins into $M_{\rm dust}$ bins, by inverting our formulation of \fdust, taken at $\Delta \rm{MS}=1$. 

\subsubsection{Accounting for the Eddington Bias} \label{sec:scatter}
Before comparing to the real data it is important to note that, while calculating the SMF-derived DMF, the Eddington bias, induced by the $f_{\rm dust}$ scatter should be taken into account.
Since we are directly employing the \citeauthor{davidzon2017} SMF, where the Eddington bias has already been corrected, using the best-fit $f_{\rm dust}$ relation to convert SMF to DMF will only reproduce the median trend and will not properly account for the full dynamic range of observed $M_{\rm dust}$.
To alleviate this, we rely on the work by \citet{loveday92}, that showed that the Eddington bias manifests itself as a Gaussian, whose width is equal to the scatter of the variable of interest, convolved to the mass function. We have thus utilised an approach similar to that used in \citet{davidzon2017} for the SMF and \citet{beeston18} for the DMF, where they successfully deconvolve their observed mass functions by using the scatter of the observed variable. As such, within each redshift bin we consider the standard deviation of the \fdust\ in a logarithmic space. We then use this scatter to create a simple Gaussian that is centred at zero, and then convolve it with our SMF-derived DMF. This allows us to better take into account the scatter of our data, and thus produce a more realistic mass function. In conclusion, we have indirectly produced two versions of SMF-derived DMFs, with and without the scatter.

\subsubsection{Comparison to the Observed Number Density}
Now we would like to compare the SMF-derived DMF to the observed number density of galaxies in the `\md-{\it{robust}}' sample. To this end, we first apply the widely used non-parametric 1/$V_{\rm max}$ method to correct for the Malmquist bias of our sample \citep{schmidt}. This method relies on assigning the $V_{\rm max}$ to each redshift bin, based on the detection limits of the survey. Effectively, this correction accounts for the fact that in a given volume, a brightness limited survey is more likely to pick up the bright sources, while missing faint galaxies, that would populate the low-$M_{\rm dust}$ end. We explicitly highlight that the $V_{\rm max}$ correction only accounts for the FIR flux rms cuts, and not the selection criteria outlined in \autoref{sec:selection}.
\par
To calculate the $V_{\rm max}$ we use the prescription from \citet{weigel2016}, which provides a volume correction for each individual source. As a first step, and for a given redshift bin, we split our sources into 0.4 dex bins in the log($M_{\rm dust}$/M$_{\odot}$) = 6 -- 11 interval. Given that the median uncertainty on the $M_{\rm dust}$ is $\sim 0.3$ dex, the following bin spacing will ensure that there is very little to no cross-contamination between mass bins. The $V_{\rm max},i$, where $i$ denotes an individual galaxy, can be then calculated as:
\begin{equation}
V_{\rm max}=\frac{A}{3}\, (d_{c}\,(z_{\rm max,i})^3- d_{c}\,(z_{\rm min,i})^3),
\end{equation}
where $d_c$ is the comoving distance and $A$ is the area, which in our case is equal to 1.38 deg$^2$.
Following \citet{weigel2016}, the $z_{\rm min,i}$ is given simply by the lower boundary of the bin. The $z_{\rm max,i}$ on the other hand can be calculated empirically, either through detection limits of individual bands or by considering a limiting mass of the survey. It however cannot exceed the maximum redshift of the bin. \par
To obtain the $z_{\rm max}$, we consider the best fit SEDs for our sources, and the rms of the parent catalogues in order to redshfit the sources to the point where they no longer fulfil our selection criteria as outlined in \autoref{sec:sample}. Using this method we however found that, for an overwhelming majority of sources, the computed $z_{\rm max,i}$ exceeds the upper boundary of the bin they belong to. Therefore, the $V_{\rm max}$ correction that we apply becomes effectively bound between the lower and the upper redshift of the bin. We find that this method works best in the lowest ($0.2 < z < 0.5$) redshift bin, with the $\langle V/V_{\rm max} \rangle$ test returning a value of 0.47. The remaining two redshift bins are significantly incomplete, with the ratio returning $0.83$ and $1.38$ respectively.  

\par
Among the other sources of incompleteness, as discussed in \autoref{sec:dmf}, here we can attempt to account for lost sources due to the sensitivity limits of the survey and failures in the deblending procedure.
 We therefore multiply our points by the loss fraction in each redshift bin that is computed as the ratio between sources in our catalogue over the sources that have \texttt{SN-IR$>1$} \footnote{\texttt{SN-IR}$^2$=$\sum_{\lambda} \, (\mathrm{S/N})^2_{\lambda}$, with $\lambda \geq 100$  $\mu$m.}
in the parent catalogue. The \texttt{SN-IR} parameter, described in greater detail in \citet{sdc1} and \citet{sdc2}, and references therein, considers a combination of FIR bands starting with 100\,$\mu$m. We thus expect that in this context, our \texttt{SN-IR} threshold can indicate whether a galaxy is intrinsically dusty.
\par

\begin{figure*}
\begin{center}
\includegraphics[width=1\textwidth]{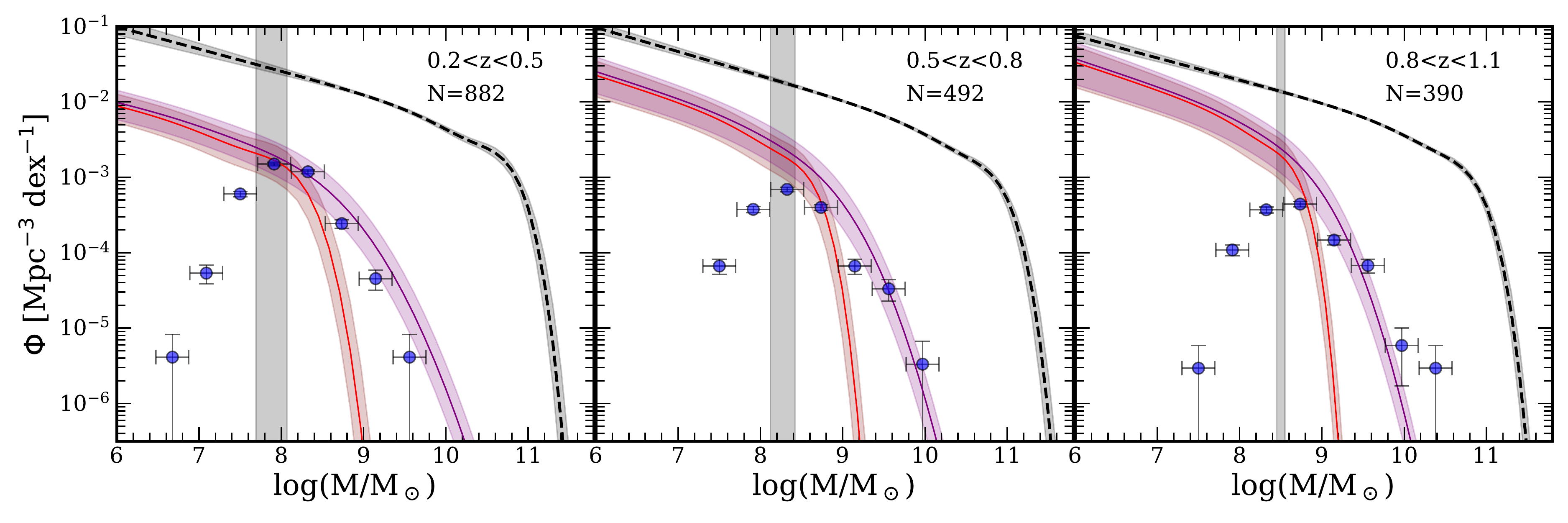}
\caption{Derived DMFs in the  $0.2<z<0.5$,  $0.5<z<0.8$, and  $0.8<z<1.1$ ranges. The dashed black line represents the original SMF for active galaxies from \citet{davidzon2017}. The purple and red lines are the DMFs obtained by converting the SMF, with and without the Eddington bias applied, respectively. The blue points are the DMF calculated directly from our data. The shaded rectangular area highlights the theoretical prediction for lim($M_{\rm dust}$) in that redshift interval.}
\label{fig:dmf2}
\end{center}
\end{figure*}

The comparison of the SMF-derived DMF with and without the Eddington bias taken into account, along with the observed volume-weighted number density of galaxies, derived as described above, for three redshift bins, is presented in \autoref{fig:dmf2}. We find that the DMF without the Eddington bias is insufficient to reproduce the observed dynamic range of $M_{\rm dust}$, particularly in the higher redshift bins, exactly as we have predicted in \autoref{sec:scatter}.  On the other hand, the SMF-derived DMF with the artificially induced bias, through the \fdust\ scatter, is in good agreement with the data in the high-mass end, further highlighting the necessity of accounting for the observational biases when inferring  relations (i.e. $M_*$ -- \md\ in our case) through the observed mass (or luminosity) functions. Although our model underpredicts the high-mass end data in the higher redshift bin, both still agree within the error bars.
At the same time though, in the low-mass regime our data significantly underestimate the number density of galaxies mirroring the incompleteness of our sample in this \md\ regime. It is worth noticing though that the turnover of the observed data perfectly coincides with the independent estimates of lim($M_{\rm dust}$), offering an indirect validation of our simulations presented in \autoref{sec:limmd} . For the analysis in the next sections, we adopt the SMF-derived DMF, which has the Eddington bias corrected out, as the final result against which we will compare previous observationally driven DMFs and theoretical predictions \footnote{Tables containing the DMFs can be accessed here: \\ \url{https://github.com/VasilyKokorev/sdc_ir_properties}.}.

\section{Gas Content of Star-Forming Galaxies} \label{sec:gas}
The inferred \md\ estimates can be used as an invaluable proxy of the gas mass (\mgas). To this end, we adopt the metallicity dependent gas-to-dust mass ratio $\delta_{\rm GDR}$ technique, that takes advantage of the relatively tight anti-correlation between the gas-phase metallicity, and the  $\delta_{\rm GDR}$ of galaxies, both in the local Universe and at high$-z$ (see e.g.\ \citealt{leroy2011,magdis2012,remy-ruyer14,genzel15}). For a source with known metallicity ($Z$) and $M_{\rm dust}$, one can estimate the amount of \mgas\  via the following relation:
\begin{equation}
M_{\rm gas} = \delta_{\rm GDR}(Z)\, M_{\rm dust},
\label{eq:deltagdr}
\end{equation}
where  \mgas\ corresponds to  $M_{\rm H_{2}}+M_{\rm HI}$, i.e. the sum of the atomic and molecular hydrogen.

Given the absence of direct metallicity measurements for our sample, we adopt the fundamental metallicity relation (FMR) of \citet{mannucci2010}. In particular, we use the \Mstar\ and SFR estimates as inputs to the FMR, and obtain metallicities calibrated for the \citet{KD02} (KD02) photoionisation models. These metallicities are  subsequently converted to the \citet{PP04} (PP04 N2) scale following \citet{kewley2008}. We then estimate  the $\delta_{\rm GDR}$ of each galaxy through the $\delta_{\rm GDR} - Z$ relation of \citet{magdis2012}, given as:
\begin{equation}
\mathrm{log}(\delta_{\rm GDR})=(10.54\pm1.00)-[12+\rm{log(O/H)]^{(0.99\pm0.12)}}
\end{equation}
\noindent and subsequently derive \mgas\ through \autoref{eq:deltagdr}, for all the sources in SDC1 and SDC2 catalogues. We propagate the uncertainties on \mgas\, by taking into account the uncertainty on $M_{\rm dust}$ and combining it with the typical scatter of 0.2 dex on the $\delta_{\rm GDR} - Z$ relation \citep{magdis12b}. These inferred $M_{\rm gas}$ estimates with associated uncertainties are included in the released catalogue.

\subsection{Gas to Stellar Mass Relation}
Similarly to \fdust, we also explore  the dependence of \fgas\  on cosmic age, $\Delta$MS and $M_*$. We utilise the same multi-parameter fitting function as before (see \autoref{eq:1}), and focus on the `\md-{\it{robust}}' sample. We calculate the Spearman rank correlation between our variables and find $M_{\rm gas}$ to be mildly correlated with log$\Delta$MS and $M_{*}$ ($\rho=0.47$ and $0.53$ respectively) and strongly negatively correlated with $t_{\rm age}$ ($\rho=-0.82$). The fitting procedure yields the following best-fit parameters:
\begin{align*}
&a_0=-0.73   &\quad a_1=1.16  \\
&b= -1.02  &\quad c_0=-0.20 \\
&c_1= 0.09  &\quad d=1.39.
\end{align*}

\begin{figure}
\begin{center}
\includegraphics[width=0.45\textwidth]{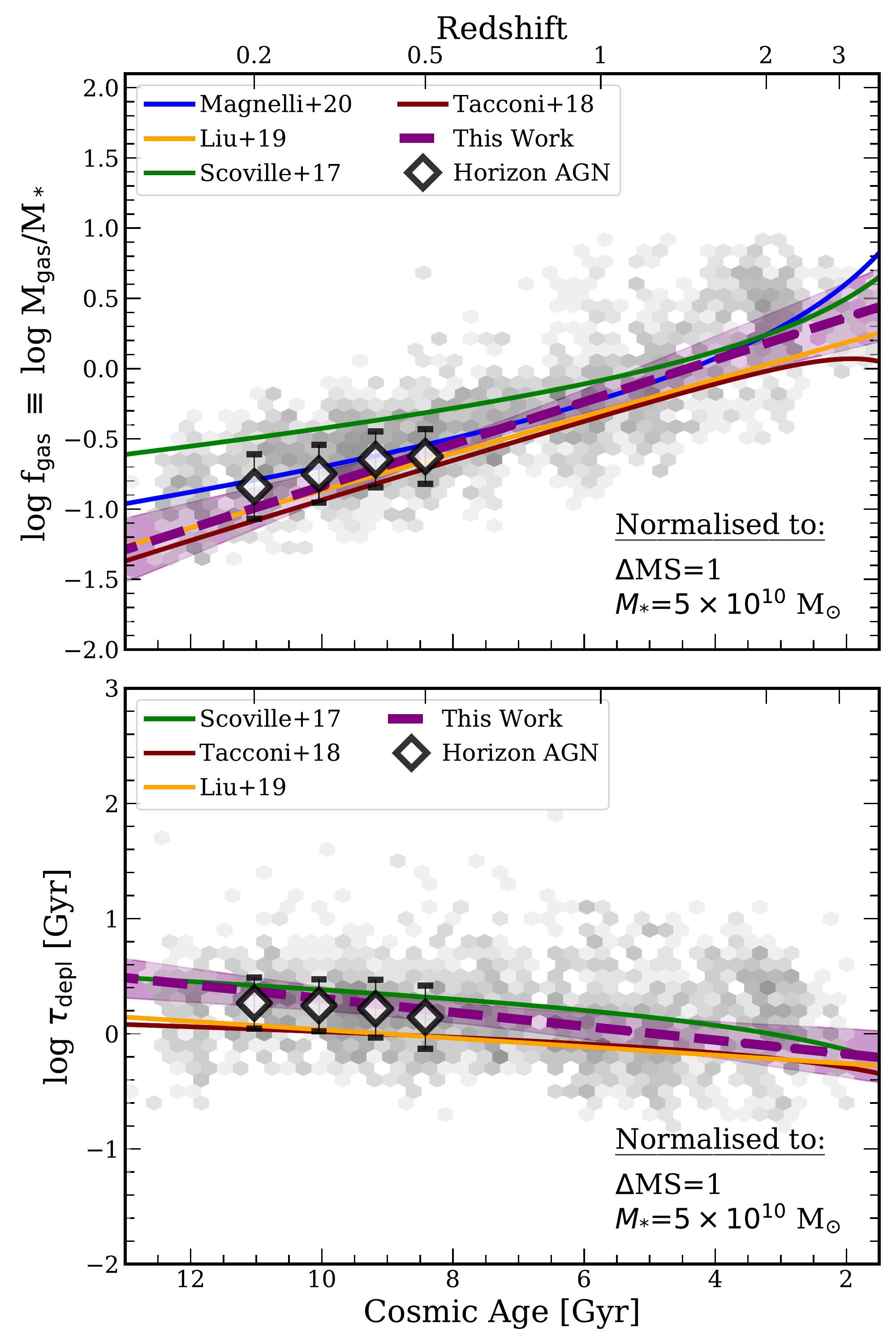}
\caption{Derived relations for $f_{\rm gas}$ (top) and $\tau_{\rm depl}$ (bottom)  as a function of $z/t_{\rm age}$. The dashed purple line shows the fit to our data, while solid coloured lines display literature results. The shaded purple region denotes the 16$^{\rm th}$ and 84$^{\rm th}$ percentile confidence intervals of our fit. The grey hexbins contain the data from `\md-{\it{robust}}' sample, and are normalised by the number count. Both the data and the derived relations have been re-scaled to $\Delta$MS=1 and $M_*=5\times10^{10}$ M$_\odot$.
White diamonds show median positions of the Horizon AGN star-forming galaxies at that redshift, normalised in the same way as our data.}
\label{fig:trends-gas}
\end{center}
\end{figure}

The best-fit \fdust-$t_{\rm age}$ (or redshift) relation along with our data, both normalised to $\rm\Delta$MS = 1 and \Mstar = 5$\times$10$^{10}$\,\msol,  are presented in \autoref{fig:trends-gas} (top). Similar to previous studies \citep{scov17,tacconi18,liu19b,magnelli20}, we find a sharp increase in the gas fraction up to $z = 2$, followed by a milder evolution at higher redshifts, a change that is noticeable only in the \fdust$-z$ parameter space. We note however that due to poor statistics and lack of spectroscopic redshifts, our data cannot reliably constrain the high-$z$ evolution of \fgas\ at  $z > 2$. We also detect a population of sources that display significantly elevated gas reservoirs for their redshift (log(\fgas)$ > 0.5$). Some of those objects have either only a $z_{\rm phot}$ estimate available, or appear to be blended in the SPIRE bands and therefore could have an erroneously large \md, and subsequently $M_{\rm gas}$ estimate assigned to them. However, among these we do identify some sources with spectroscopic redshifts, that are also `clean'/isolated in the IR maps. In particular we find $\sim 40$ such sources with log(\fgas)$ > 0.5$) and six with log(\fgas)$ >1$ that, as we discuss later, we coin as `gas-giants'.

\subsection{Evolution of Depletion Time}
Finally, we focus our attention on the depletion time $\tau_{\rm depl}=M_{\rm gas}/\mathrm{SFR}=1/\mathrm{SFE}$. We employ the same fitting technique as before, and explore the evolution of $\tau_{\rm depl}$ with cosmic age, $\Delta$MS and $M_*$
in a multi-dimensional parameter space, for the `\md-{\it{robust}}' sample. The $\tau_{\rm depl}$ correlates mildly with 
age and $\Delta$MS (Spearman $\rho= -0.46$ and $-0.58$, respectively) and weakly with $M_*$ ($\rho=-0.23$). The best fit parameters are as follows:
\begin{align*}
&a_0=-1.65   &\quad a_1=1.42  \\
&b= -0.95  &\quad c_0=-0.02 \\
&c_1= 0.12  &\quad d=0.37.
\end{align*}
We show the best-fit $\tau_{\rm depl}$-$t_{\rm age}$ relation along with our data, both normalised to $\rm\Delta$MS = 1 and \Mstar = 5$\times$10$^{10}$\,\msol,  are presented in \autoref{fig:trends-gas} (bottom). In line with the previous studies of $\tau_{\rm depl}$ by \citealt{scov17,tacconi18,liu19b}, we recover a relatively weak decrease of depletion time (or increase in SFE) with redshift.

\section{Discussion} \label{sec:disc}
\subsection{On the Dust and Gas Scaling Relations}

The recovered trends between \md, \Mstar\ and \mgas\ and their evolution with redshift offer a test-bed against which  theoretical and previous observationally driven studies can be compared to. As shown in  \autoref{fig:trends-dust} and \autoref{fig:trends-gas}, our analysis yields  \fdust\ and \fgas\ evolutionary tracks consistent with those presented in  \citet{tacconi18,liu19b} and to a smaller degree to those reported in \citet{scov17,magnelli20}. As discussed earlier, the mild tension between the latter works and our results could be primarily attributed to the choice of the fitting function.\par
The $f_{\rm dust}$ and $f_{\rm gas}$ in our sample of SFGs  increase rapidly from $z=0$ to $z=1$,  peak around $z\sim2-3$, and then remain roughly constant. It is however a point of contention whether the latter is driven by actual physical processes or is a consequence of the scarcity of data at $z>2$.  It it also worth mentioning that our analysis points towards a milder evolution of \fdust\ ($-0.8$\,dex)  compared to \fgas\ ($-1.3$\,dex) from $z=2$ to $z=0$ with the latter dropping $\sim 3 \times$ faster. This is aligned with the evolution of $\rho_{\rm dust}$ and $\rho_{\rm gas}$ derived by the ALMA stacking analysis of  \citet{magnelli20}, and could in fact reflect the evolution of metallicity, and thus of $\delta_{\rm GDR}$, for fixed \Mstar\ towards lower redshifts. \par
At the same time, the decrease of \md\ with decreasing redshift, for fixed \Mstar,  can be attributed  to either the destruction of dust grains by interstellar radiation fields, or their incorporation into the stellar population. This is discussed in more detail  in \citet{donevski20}, where they also report a decreasing  \fdust\ from earlier cosmic age to the present epoch. We note that the observed trend could also mirror the overall decline in the SFRD in the Universe from $z=2$ to the present day, that points towards lower star-formation activity  and thus lower dust production at later cosmic epochs. Finally, at a fixed redshift, both \fdust\ and \fgas\ decrease as a function of $M_*$ (as indicated by a negative value of the fitting parameter $b$, see \autoref{eq:1}), in line with previous studies (e.g.\ \citealt{magdis2012,magnelli20} and references therein).\par
In addition to observational studies, we can also compare our results to theoretical predictions. To this end, we consider the  Horizon AGN (HAGN) hydrodynamical simulations in the COSMOS field  \citep{dubois14,laigle19} and draw a sample of SFGs ($\Delta$MS$>$0.3) in the $z=0.2-0.5$ range, selected to meet the $M_*$ completeness of the COSMOS 2015 survey and which fall within a simulation box of 143 Mpc per side \citep{dubois14}. To measure $\Delta$MS for each galaxy we considered the \Mstar\ and the 100 Myr averaged SFR from the simulations. Also, since $M_{\rm dust}$ is not an explicit paramater of HAGN galaxies, we used a constant $\delta_{\rm GDR} = 100$ to convert the \mgas\ values, as derived from the simulations, to \md. We then use the \mgas, \Mstar, and \md\ of the simulated galaxies to infer \fgas\ and \fdust. The median values and their scatter, re-normalised to MS ($\Delta$MS = 1) and  $M_*=5\times10^{10}$ M$_{\odot}$ in four redshift bins, are presented and compared to the real data in \autoref{fig:trends-dust} and \autoref{fig:trends-gas}.
We find a good agreement between the theoretical predictions and our observationally driven trends (in the $<0.2<z<0.5$ range at least) indicating that the HAGN simulation can successfully reproduce the baryonic components of the galaxies and its evolution with redshift. Conversely, the agreement of our results with both theoretical and observational studies, provides an extra indirect validation for the performance of our new SED fitting code.

\subsection{On the Evolution of Depletion Time}
As with \fdust\ and $f_{\rm gas}$, our recovered trends, that connect $\tau_{\rm depl}$ to redshift, $\Delta$MS and $M_*$, show similar behaviour to the ones presented in \citealt{scov17}, and to a lesser extent \citealt{tacconi18,liu19b}. The dependence of $\tau_{\rm depl}$ on $M_*$ is relatively weak across all studies, however, similarly to \citeauthor{liu19b}, we find that the depletion time for high-mass galaxies increases from early cosmic ages towards present times, while low-mass galaxies display an opposite trend of decreasing $\tau_{\rm depl}$ with cosmic age. As discussed in \citet{liu19b} and \citet{hodge20}, this could be a signature of downsizing, meaning that more massive galaxies evolve at earlier times. \par
During our analysis we find $\tau_{\rm depl} \sim (1+z)^{-1.07}$, which is more reflective of the scaling relations derived in \citet{scov17} ($\tau_{\rm depl} \sim (1+z)^{-1.04}$), rather than weaker dependencies ($\tau_{\rm depl} \sim (1+z)^{-0.62}$ and $ (1+z)^{-0.58}$) found by \citet{tacconi18} and \citet{liu19b} respectively. As expected, and in line with the literature results, we also find that galaxies above the MS (at a fixed $M_*$ and $z$), form stars with a much higher efficiency (lower $\tau_{\rm depl}$), than their MS counterparts, with $\tau_{\rm depl} \sim \Delta \mathrm{MS}^{-1.68}$. We would also like to caution the reader and highlight the fact that \citeauthor{tacconi18} and \citeauthor{scov17} use functional forms that are different from ours, when fitting for evolution of $\tau_{\rm depl}$. For example, \citeauthor{tacconi18} consider additional dependence on the effective radius $R_{\rm e}$, which might inadvertently carry some redshift dependence. As such the fitted exponents are not necessarily directly comparable.
The differences between evolutionary trends could also be attributed to the different samples used (see e.g. \citealt{hodge20}). 
\par
Presumably, the existence of these outliers can be explained by an increased SFE, which results from major-merger events (see e.g. \citealt{scov17,cibinel19}). In fact, galaxies that lie above the MS are also found to have increased gas fractions \citep{dekel09,tacconi20}, which is attributed to a more efficient gas accretion from the cosmic web, but the enhanced gas reservoirs are still not large enough to explain significantly enhanced sSFR. The debates regarding the exact reason, which results in an onset of a SB - like mode of star-formation, are still ongoing, however it seems very likely that it is a combination of both increased gas fractions and enhanced SFE. We find that our sample supports this notion, with galaxies above the MS having both large gas reservoirs with median log($f_{\rm gas}$)=0.15, meaning that gas mass reservoirs take up $\sim 59 \%$ of the total baryonic matter, and also relatively short depletion times of $\sim 400$ Myr. Our $M_{\rm gas}$ values were however derived with a general FMR, assuming solar-like metallicities. This, however, might not be applicable for SBs, which can display elevated metallicities due to the increased sSFR. In fact it has been shown (see e.g. \citealt{silverman15}), if SBs had super-solar metallicites, it would drive down $\delta_{\rm GDR}$ together with $f_{\rm gas}$, and in turn result in increased SFE, thus implying that only the SFE is responsible for galaxies being elevated above the MS.

\subsection{On the DMFs and the Theoretical Predictions}
\begin{figure}
\begin{center}
\includegraphics[width=0.5\textwidth]{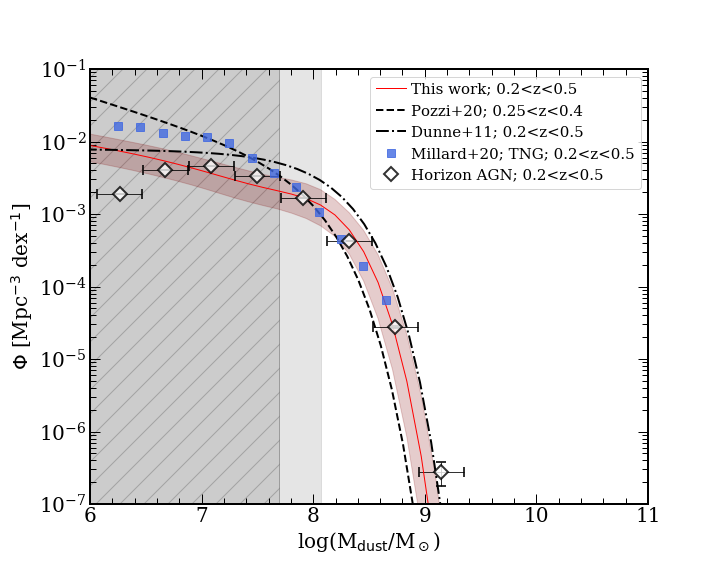}
\caption{A compilation of the theoretical and observationally derived DMFs in the  $0.2<z<0.5$ range. The \citet{davidzon2017} SMF, converted to DMF, is shown as the solid red line, with the shaded area corresponding to the $1\sigma$ uncertainty. The dashed and dash-dotted black lines correspond to the DMFs of  \citet{pozzi20} and \citet{dunne11} respectively, rescaled to $\kappa_{250 \mu \mathrm{m}}=0.51$ m$^2$ kg$^{-1}$. The white diamonds and the blue squares depict the theoretical predictions of the HAGN and TNG simulations from  \citet{dubois14} and \citet{millard20} respectively. The grey shaded region highlights the \md\ regime below the lim($M_{\rm dust}$) of our sample, as derived in \autoref{fig:lim_md}. The hatched region denotes the \md\ regime where our sample becomes severely limited, i.e. $>1\sigma$ below lim($M_{\rm dust}$).}
\label{fig:dmf_comparison}
\end{center}
\end{figure}

With the derived DMF in hand, we are also in position to bring together our findings with those presented in previous observationally driven studies and provide a direct comparison to the theoretical predictions as inferred by recent simulations. For our purposes, we focus on the  $0.2 < z < 0.5$ redshift interval that contains the majority of our objects and offers the most robust statistical analysis. These results are shown in \autoref{fig:dmf_comparison}. 
\par

We first compare our DMF to that presented in \citet{pozzi20}, based on a PACS-$160\,\mu$m selected sample of SFG. In \autoref{fig:dmf_comparison}, the two DMFs appear to be in tension both in the high and the low-mass end, with our compilation over-predicting the number density of galaxies with high \md\ and under-predicting that of less dusty sources. The discrepancy between the two DMFs can be attributed to the choice of fitting methods/templates, the adopted $\kappa_\nu$ to infer \md\ as well as to selection effects. For example, the DL07 templates adopt a $\kappa_{250 \mu \mathrm{m}}$ of 0.51 m$^2$ kg$^{-1}$, while the analysis in \citeauthor{pozzi20} uses $\kappa_{250 \mu \mathrm{m}}=0.4$ m$^2$ kg$^{-1}$, which would result in a 0.1 dex smaller $M_{\rm dust}$ estimates.
It is also important to point out that \citeauthor{pozzi20} compared their MBB SED fitting to \texttt{MAGPHYS}, finding that their MBB method recovers systematically lower $M_{\rm dust}$. Indeed, choice of the fitting methodology can induce up to a factor of two difference in derived $M_{\rm dust}$ (see e.g. \citealt{magdis13} and an in-depth comparison in \citealt{berta16}). Moreover, for a flux limited survey, the mere selection at $\lambda_{\rm obs}$=160$\,\mu$m could introduce a bias towards warmer sources that for fixed \lir\ have lower \md\ and which could explain the small number density of sources with log(\md/\msol) $>$ 8.  While it is not possible to correct for the effects of selection and broader SED fitting methodology, we have rescaled the \citeauthor{pozzi20} DMF to have the same $\kappa_{250 \mu \mathrm{m}}$ as was adopted in our analysis.

We also compare our results to \citet{dunne11}, that computed a DMF based on a sample of 250\,$\mu$m selected galaxies. 
For our comparison, we have rescaled their DMF by $-0.24$ dex, to account for the difference in $\kappa_\nu$.
Contrary to \citet{pozzi20}, we now find that \citet{dunne11} over-predicts the number density of dusty galaxies at high dust masses. This again can be understood in terms of selection effects since the 250\,$\mu$m selection could bias the sample towards cold sources and thus to higher \md\ values (again, for a flux limited survey). While our criterion for at least one detection at $\lambda_{\rm rest}>150\,\mu$m could be perceived as similar to a 250\,$\mu$m selection at $z\sim0.3$, we note that the requirement for two extra  detections at $\lambda_{\rm rest} < 150\,\mu$m, and the super-deblendend catalogues, that allow for the detection of fainter than the nominal confusion noise in the SPIRE bands, ease any bias towards either intrinsically cold or warm objects. This is further supported by the fact that our SMF-derived DMF, where the Eddington bias has already been corrected, falls directly between the calculations from \citeauthor{pozzi20} and \citeauthor{dunne11} (\autoref{fig:dmf_comparison}). In conjunction with the derivation of a \avu\ -- $z$ relation that is in excellent agreement with the stacking analysis of  \citet{bethermin15}, this suggests that the careful treatment of selection criteria and of the detection limits of our parent sample has allowed us to gain 
a unique and unbiased perspective on the evolution of dust properties of COSMOS galaxies.
\par

Finally, we compliment our analysis by comparing our DMF to the theoretical predictions of the HAGN and   IllustrisTNG simulations \citep{millard20}. In order to produce a HAGN DMF, we define a simulated sample following the procedure described
above, bin the galaxies in 0.4\,dex intervals of \md, and normalise by the volume of the simulation ($4\times(142)^3$ Mpc$^{3}$). For the IllustrisTNG simulation, \citet{millard20} consider multiple TNG100 snapshots in a box size of 106 Mpc per side, comparable to the HAGN simulated subset presented earlier. The TNG-DMF is constructed through the \md\ values of the simulated galaxies, derived in post-processing through a fixed dust-to-metals ratio of 0.5.\par

Unlike real data, simulations do not suffer from observational bias, and as such should be compared to DMF derived from the SMF, without adding the Eddington bias. As shown in \autoref{fig:dmf_comparison}, both HAGN and IllustrisTNG  are in excellent agreement with
the high-mass end of our SMF-derived DMF. Notably, the HAGN-DMF is also consistent with our results at the low-mass end down to  $M_{\rm dust}\sim 10^{7}$ M$_{\odot}$. We recall, that for simplicity, when converting $M_{\rm gas}$ to $M_{\rm dust}$ for the HAGN sample, we considered a universal $\delta_{\rm GDR}=100$. However, for sources with lower \Mstar\ ($<10^{8}$ $M_{\odot}$)  and thus with  sub-solar metallicities, a larger $\delta_{\rm DGR}$ ($\sim 150$) is probably more applicable \citep{remy-ruyer14}. This would translate into a $\times1.5$ downward correction for the low-mass HAGN bins, bringing them into exact agreement with our DMF down to $M_{\rm dust}\approx 10^{7}$ M$_{\odot}$. We note that this \md, assuming an average $M_{\rm gas}$/\md $\approx$ 100, corresponds to the \Mstar\ completeness limit of the simulation ($M_*\approx 10^{9}$ M$_{\odot}$). Therefore, the observed decline of the number density of the HAGN galaxies at $M_{\rm dust} \leq 10^{7}$ M$_{\odot}$ is fully consistent with the expectations.

In comparison to the TNG-DMF though, we predict a factor of $\times 2.5$ fewer objects at the low-mass end. This tension could arise from the incompleteness of our sample at the low-mass end, that leaves the slope of the $M_{\rm dust} - M_{*}$ relation at \md\ $<$ 5 $\times$ 10$^{7}$ \msol\ largely unconstrained. We are thus unable to ascertain whether this discrepancy is caused by the limitations of our sample, or whether the TNG simulations over-predict the number density of the galaxies in the low-mass end.  

Put together, these comparisons indicate that, at least down to \md\ $\approx$ 5 $\times$ 10$^{7}$ \msol, our \md$-$\Mstar$-z$ relation and the resulting DMFs are robust and fully consistent with the theoretical expectations. 
\par

\subsection{Population of Gas Giants}
\begin{figure*}
\begin{center}
\includegraphics[width=1.\textwidth]{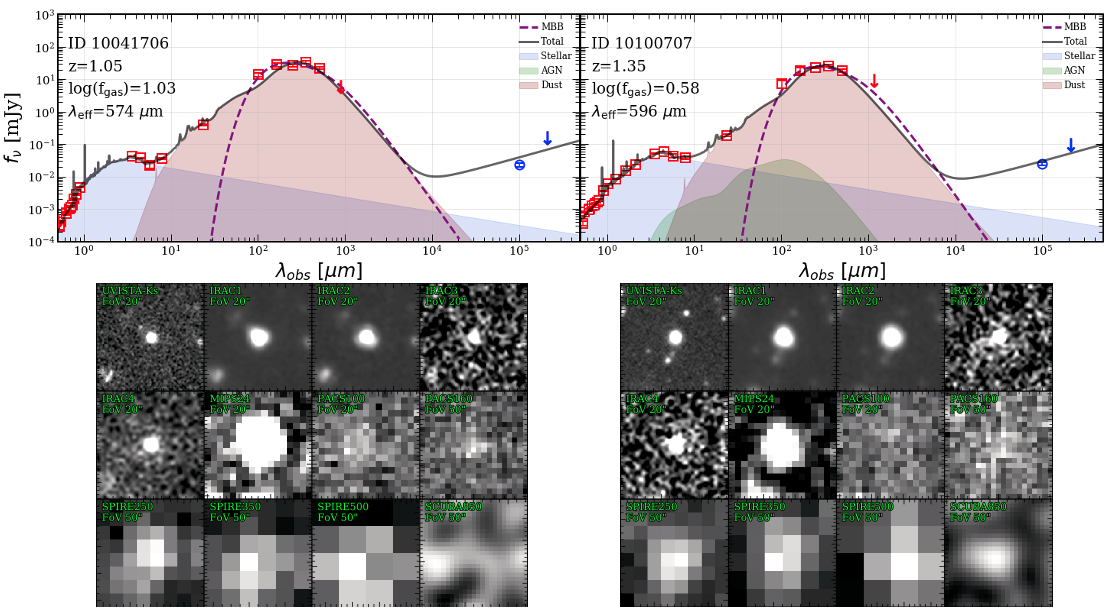}
\caption{\textbf{Top:}. Photometry and best-fit SEDs for two `gas-giants' (log(\fgas) $>$ 0.5) at $z_{\rm spec}$ = 1.05 and $z_{\rm spec}$ = 1.35. Colour coding and symbols are the same as in \autoref{fig:seds}, with the addition of a dashed purple line that shows the best-fit optically thick MBB. The $\lambda_{\rm eff}$ (in rest-frame) at which the SED becomes optically thick ($\tau=1$) is displayed in the panels. \textbf{Bottom:} NIR-FIR cutouts of these objects. The cutout sizes range from $20''$ in the NIR-MIR range to $50''$ in FIR.}
\label{fig:gas_giant}
\end{center}
\end{figure*}

As briefly discussed in \autoref{sec:gas}, during our analysis we identified  some extreme outliers from the average \fdust\ and \fgas\ evolutionary trends (\autoref{fig:trends-dust} and \autoref{fig:trends-gas}), and which typically have log(\fgas) $>$  0.5, i.e. their gas mass reservoir takes $\sim75\%$ of their baryonic matter. Since $z_{phot}$ could be a major source of uncertainty in both \Mstar\ and \mgas,  before looking further into this population of `gas-giants', we first narrow down our sample to  spectroscopically confirmed sources. We then examined the individual SEDs and the cut-out images of the remaining sources in order to identify either poor coverage of the FIR peak or blending issues that could result in erroneously large \mgas\ estimates. With the above considerations, we are left with 41 objects whose extreme \fgas\ can only be explained by gigantic \mgas\ reservoirs. This population spans a wide range in redshift ($0.21<z<4.05$, $\langle z \rangle$ = 1.34), with $9.0<\mathrm{log}(M_*/\mathrm{M}_\odot)<11.3$, $\langle \mathrm{log}(M_*/\mathrm{M}_\odot) \rangle=10.3$ and $0.11<\Delta$MS$<14.2$, $\langle \Delta$MS$\rangle = 1.8$. The best-fit SEDs of two such objects are presented in \autoref{fig:gas_giant}. We also note that these two sources are otherwise unremarkable, and have what can be considered `typical' values for the log($M_*)\sim 10.7$, and also do not appear to be strong SBs ($\Delta$MS=3.8 and 2.2, respectively for 10041706 and 10100707). Furthermore, the cutouts presented in \autoref{fig:gas_giant} indicate that these sources do not appear to be blended, therefore the only unusual characteristic that they possess, seems to be an elevated $M_{\rm gas}$.

\begin{figure}
\begin{center}
\includegraphics[width=0.45\textwidth]{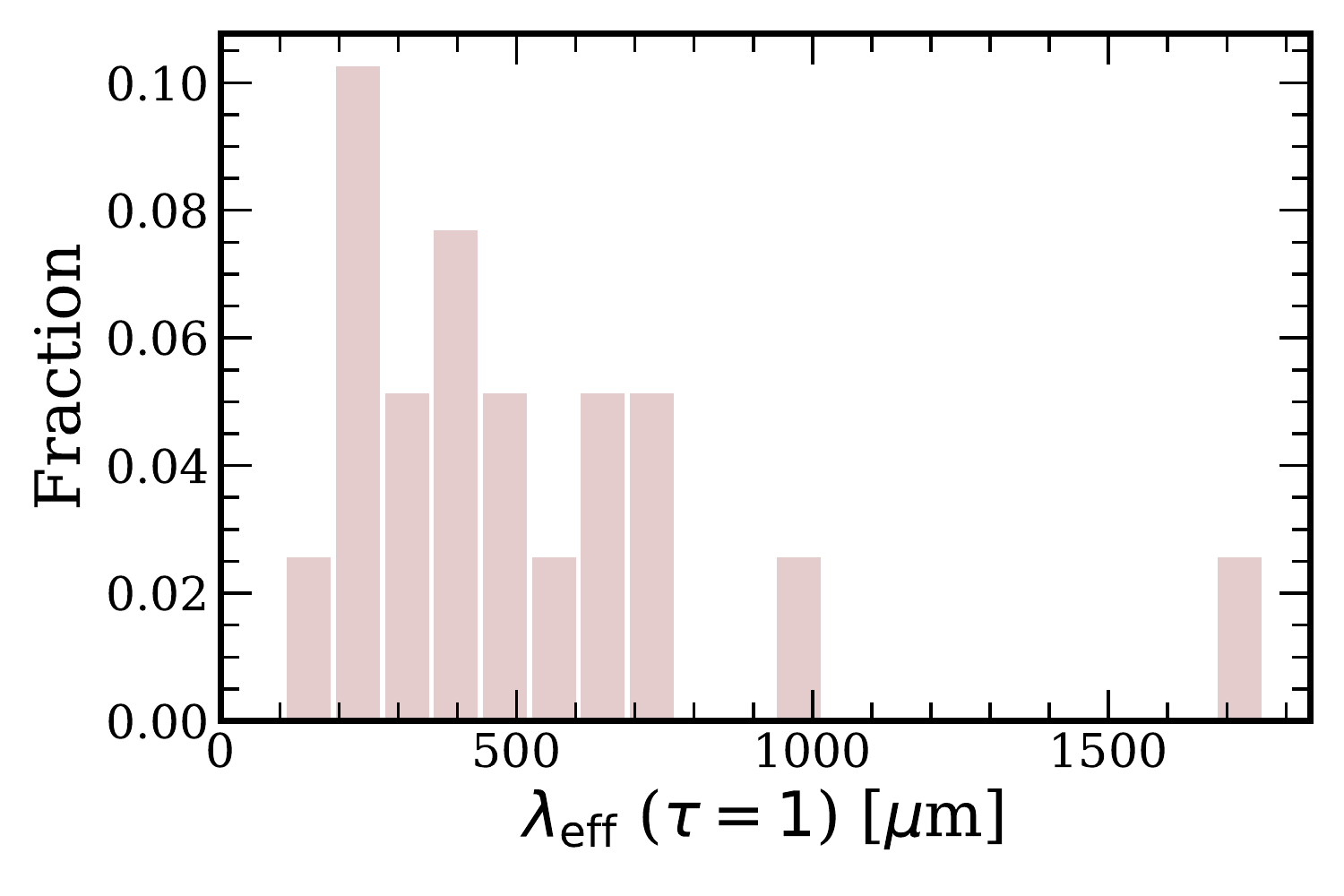}
\caption{Distribution of $\lambda_{\rm eff}$, below which the emission of the `gas-giants' in our sample becomes optically thick ($\tau=1$), as inferred by MBB models of general opacity.}
\label{fig:leff}
\end{center}
\end{figure}

A possible explanation for the very high \md\  and subsequently \mgas\ estimates for these galaxies, other than an extremely low $\delta_{\rm GDR}$, could be an optically thick FIR emission. In this scenario, the attenuation of the emission in the Wien part of the spectrum makes the galaxy \textit{appear} cold, leading to an overestimate of its true \md\ \citep[e.g.][]{jin19,cortzen20}. Since the DL07 models assume that the galaxy is optically thin at $\lambda_{\rm rest} > 1$\,$\mu$m, to test this scenario we employed Modified Black Body (MBB) models of general opacity, leaving the effective wavelength ($\lambda_{\rm eff}$) at which $\tau=1$ as a free parameter (e.g. see \citealt{casey12} for the functional form). We fixed the Rayleigh-Jeans slope to $\beta=1.8$, and only fit the available photometry of each source at $\lambda_{\rm rest} >40\,\mu$m. Due to the large number of free parameters in this model, we further limit our sample to sources with five or more IR detections. Out of the 41 `gas giants', we are thus able to constrain $\lambda_{\rm eff}$ for only 19. The distribution of the inferred $\lambda_{\rm eff}$ values is presented in \autoref{fig:leff}. 

We find that the vast majority of these objects have $\lambda_{\rm eff} >100\,\mu$m, which implies that these galaxies could be optically thick in the FIR. The unusually high \fdust\ and \fgas, can therefore be incorrect simply due to the optical depth effects. Indeed, a comparison between the inferred \md\ estimates for an optically thin and optically thick case, yields an average ratio  of $\sim \times1.8$ for our sample. However, while this correction would  reduce the average \fgas\ of the `gas-giants' from $\langle$log(\fgas)$\rangle$ = 0.72 to $\langle$log(\fgas)$\rangle$ = 0.47, this is still substantially larger with respect to the average population of SFGs. Finally, the real \fgas\ could in fact be lower if the $M_*$ of the sources is underestimated, a scenario that is indeed in line with a dusty, optically thick ISM.

To understand whether these objects do indeed host unusually high gas mass reservoirs and shed light on their nature, additional observations, with either ALMA or NOEMA, of  $M_{\rm gas}$ tracers (e.g.\ CO and CI) are necessary.

\section{Summary} \label{sec:conc}
In this work we present an in-depth analysis of the evolution of the FIR properties of SFGs by studying a large sample of sources drawn from the publicly available infrared catalogues in the GOODS-N and COSMOS fields \citep{sdc1,sdc2}. Both catalogues are contructed based on a novel `Super-Deblending' technique that allows prior-based photometry in the highly confused \emph{Herschel} and \emph{SCUBA+AzTEC} maps. 

In the process, we developed a new panchromatic SED fitting algorithm - \texttt{Stardust} - to fit a linear combination of stellar, AGN and infrared (star-forming) templates, in an attempt to perform a coherent, systematic and homogeneous analysis in the two fields. Our fitting tool has two key advantages. Firstly, the best-fit model is a set of coefficients rather than a single template, thus it does not rely on iterating through thousands of possible template combinations, speeding up the fitting process by a factor of $\sim 10$ compared to other multi-wavelength fitting  available codes. Secondly, the fitting process does not impose energy balance between absorption in the  UV/optical and emission in the IR, treating the stellar and the dust emission components independently. As such, it is very relevant for sources where the stellar and dust emission are not co-spatial. The code itself is also highly modular, allows for user input templates, and is publicly available.

A first product of this new software is a multi-parameter catalogue that contains the FIR properties of $\sim5,000$ infrared bright galaxies in GOODS-N and COSMOS. The extracted parameters, their uncertainties and the matched photometry from the original `Super-Deblended' catalogues are released and are also publicly available \footnote{Tables containing the DMFs can be accessed here: \\ \url{https://github.com/VasilyKokorev/sdc_ir_properties}}. The list of output best-fit parameters and the structure of the released catalogue can be found in \autoref{tab:catalogue}.

We subsequently used the extracted parameters to explore the evolution of the FIR properties of SFGs and recover scaling relations, aided by a careful set of simulations that quantify the underlying selection effects and biases of our sample in terms of limiting \md, \lir\ and \avu. In particular, we parametrised the \fdust\ of galaxies as a function of their cosmic age, $M_*$, and $\Delta$MS. The median \fdust\ is found to increase by a factor of $\times 10$ from $z = 0$ to $z = 2$ with a mild, if any, evolution at higher $z$. Through the metallicity dependent $\delta_{\rm GDR}$ technique, we also derive the evolution of \fgas\ and find it to be consistent with  previous observational studies, as well as with theoretical predictions. 

Furthermore, we constructed the DMF up to $z = 1$ by converting the SMF of SFGs to a DMF, through the evolution of \fdust\ and its scatter, as parametrised in our study. A comparison of the derived DMFs to the theoretical predictions of the HAGN and TNG100 simulations in the $0 < z < 0.5$ range reveals an excellent agreement down to a limiting \md$\sim 5\times10^{7}$\,\msol, where due to poor statistics we cannot adequately constrain the \md-\Mstar\ relation. 

Finally, we identified a population of SFGs with extreme log(\fgas) $>$ 0.5, that we coin as `gas giants'. The \fgas\ excess of these galaxies compared to the average SFG  population persists even when opacity effects in the FIR emission are taken into account. Follow-up observations targeting alternative \mgas\ tracers are necessary to confirm the extreme nature of these systems.


Some further remarks that we would like to emphasise: 
\begin{itemize}
    \item The effect of the photo$-z$ uncertainty in the derivation of \md (and \lir) is not negligible and should be accounted for. We find that a photo$-z$ uncertainty of $\Delta z$/(1+z$_{\rm spec}$) $\sim 0.02$, characteristic of fields like GOODS-N and COSMOS, introduces an extra $20 \%$ of scatter in the derivation of \md\ and \lir.
    \item As already discussed in the literature, the uncertainty in the derivation of \md\ increases substantially in the absence of a data point in the R-J tail ($\lambda_{\rm rest} > 150\,\mu$m). However, the presence of three data points in the mid-to-FIR could securely constrain \md\ within a factor of $\sim 0.3$ dex, even if the last available data point is at $\lambda_{\rm rest} \approx 150\,\mu$m.   
    \item When using the \md\ -- \Mstar\ -- $z$ scaling relations to convert SMF to DMF (or similarly to gas mass functions), the scatter of the relations used for the transformation should be taken into account for a proper comparison to the data. Similarly, any attempt to derive scaling relations between two (or more) parameters through the comparison of mass (or luminosity) functions inferred through the modelling of the observed number densities should entail a proper consideration of the scatter of the parameters in question. 
    \item Both the warm $M_{\rm dust}$ and the warm IR emission arising from the PDRs are increasing with respect to the cold $M_{\rm dust}$ and cold dust emission as we move above the MS, indicative of more compact/active star-forming activity. Subsequently, the clear and relatively tight trend of decreasing $M^{\rm cold}_{\rm dust}$ (for fixed \lir) with $\rm \Delta$MS is less pronounced for $M^{\rm warm}_{\rm dust}$. This enforces the overall picture where SBs are characterised by higher star-formation efficiencies and with a larger fraction of their \md\ being exposed to more intense radiation fields.
    
\end{itemize}

\acknowledgments
We would like to thank the anonymous referee for a number of constructive suggestions. We thank Yohan Dubois and Clotilde Laigle for providing us with HAGN data and their input towards the discussion of simulations, Jenifer Millard for providing us with the DMF data, and Benjamin Magnelli for the help in understanding the dust mass density evolution. The Cosmic Dawn Center is funded by the Danish National Research Foundation under grant No. 140. GEM acknowl- edges the Villum Fonden research grant 13160 “Gas to stars, stars to dust: tracing star formation across cosmic time,” grant 37440, “The Hidden Cosmos,” and the Cosmic Dawn Center of Excellence funded by the Danish National Research Foundation under the grant No. 140. FV acknowledges support from the Carlsberg Foundation research grant CF18-0388 “Galaxies: Rise And Death”. ID acknowledges support from the European Union’s Horizon 2020 research and innovation programme under the Marie Sk\l{}odowska-Curie grant agreement No. 896225. ST, GB and JW  acknowledge support from the European Research Council (ERC) Consolidator   Grant funding scheme (project  ConTExt,  grant  No.   648179). SJ acknowledges financial support from the Spanish Ministry of Science, Innovation and Universities (MICIU) under AYA2017-84061-P, co-financed by FEDER (European Regional Development Funds). DL acknowledges funding from the European Research Council (ERC) under the European Union's Horizon 2020 research and innovation program (grant agreement No. 694343). This work has made use of the CANDIDE Cluster at the Institut d'Astrophysique de Paris and made possible by grants from the PNCG and the DIM-ACAV.

%

\vspace{5mm}



\clearpage
\appendix 
\section{SED Fitting} \label{sec:appendix}
\subsection{Draine \& Li (2007) Templates}
In our fitting routine we utilise the dust models of \citet{dl07}, with the updated opacity from \citealt{draine14}.
These models aim for a robust and physically motivated approach to SED fitting in both MIR and FIR, as well as allow us to calculate the amount of luminous dust. \par
The description of the dust locked in the interstellar medium is one of a mixture of carbonaceous and amorphous silicate grains, with their sizes and distributions following the extinction law in the Milky Way, Large Magellanic Cloud (LMC) and the bar region of the Small Magellanic Cloud (SMC). The carbonaceous grains behave similarly to the polycyclic aromatic hydrocarbon (PAH) molecules, with their properties given by the PAH index $q_{\rm PAH}$, that is defined as a fraction of dust mass locked into the PAH grains. \par
The models provide a bimodal description of the environments containing the interstellar dust: the diffuse ISM and the PDRs. The bulk of the dust mass is thought to be located in the cold and diffuse part of the interstellar medium, that is being heated by a radiation field of a constant intensity $U_{\rm min}$. A smaller proportion of the mass budget described by the $\gamma$ index is exposed to a gradient of radiation intensities ranging from $U_{\rm min}$ to $U_{\rm max}$, and is supposedly located in the warmer PDRs. Although these warm regions normally contain only a small fraction of the total dust mass, they can make a substantial contribution to the luminosity in the mid-IR SEDs. As described by DL07, the infinitesimal proportion of $dM_{\rm dust}$ exposed to radiation fields between $U$ and $U+dU$ can be modelled by a power law distribution, and in the case of the diffuse ISM where $U_{\rm min}=U_{\rm max}$, by a Kronecker $\delta$-function. This leads to the following description:
\begin{multline}
dM_{\rm dust}=(1-\gamma)\,\delta(U-U_{\rm min}) \\ 
+\gamma \, M_{\rm dust}\frac{\alpha-1}{U_{\rm min}^{1-\alpha}-U_{\rm max}^{1-\alpha}}
\end{multline}
with $U_{\rm min}\leq U_{\rm max}$ and $\alpha \neq 1$. The parameter - $\gamma$ - is the fraction of dust mass locked into the high starlight intensity regions described by the power-law, $\alpha$ gives the distribution of radiation intensities in the PDRs and $M_{\rm dust}$ as the total dust mass. \par
The methods described in DL07 allow us to compute a distribution of temperatures for all particles: ones that are small so their size makes them susceptible to the effects of quantised heating, and the larger ones where the steady-state temperatures dictated by the stellar radiation and radiative cooling equilibrium take hold. One can compute an averaged IR emission for a given grain type by first considering their temperature distribution and cross sections, and then sum everything up to obtain the specific mass weighted power that is being radiated by the dust exposed to starlight of intensity $U$. By integrating these numerical recipes from $U_{\rm min}$ to $U_{\rm max}$, one can obtain the power per unit frequency per unit mass $p_\nu (q_{\rm PAH},U_{\rm min},U_{\rm max},\alpha)$. \par
In line with DL07, one can then model the galaxy spectral energy distribution as a linear combination of the diffuse ISM and the PDRs. This can be written as follows: 
\begin{equation}
j_{\nu}=(1-\gamma) \, j_{\nu}[U_{\rm min},U_{\rm max}]+j_{\nu}[U_{\rm min},U_{\rm max},\alpha]
\end{equation}
where $j_{\nu}$ is the emissivity per hydrogen nucleon. If one now considers a galaxy at some distance $D_{\rm L}(z)$, the received flux density can be written as:
\begin{equation}
f_{\nu}=\frac{M_{\rm H}}{m_{\rm H}}\frac{j_{\nu}}{D^2_{\rm L}(z)}.
\end{equation}
Since $j_{\nu}$ is the quantity contained in the DL07 models, the normalisation extracted from the fitting represents the total number of hydrogen nucleons, and can be then converted to the luminous dust mass.\par
The total luminosity contained in both dust components can be written as:
\begin{equation}
L_{\rm dust}=\langle U \rangle P_{0}M_{\rm dust}
\label{eq:4}
\end{equation}
with $\langle U \rangle$ representing a mean intensity of the radiation field, given by:
\begin{equation} \label{eq:5}
\langle U\rangle=(1-\gamma)\,U_{\min}+\frac{\gamma\, \mathrm{ln}\, (U_{\rm max}/U_{\rm min})}{U_{\rm min}^{-1}-U_{\rm max}^{-1}},
\end{equation}

for $\alpha=2$ and $P_{0}$ denoting the power absorbed per unit dust mass in a radiation field of intensity $U=1$. \par
In principle, one could think of these models as having six effective free parameters - $q_{\rm PAH}$, $U_{\rm min}$, $U_{\rm max}$, $\alpha$, $\gamma$ and \md - acting as the normalisation as described above. It has been shown in \citet{draine07} that the parameter space is insensitive to the adopted dust model (MW, LMC and SMC) and the values of $\alpha$ and $U_{\rm max}$. It is possible thus to recover a wide range of properties of various SEDs by fixing these to $\alpha=2$ and $U_{\rm max}=10^6$. The values of $U_{\rm min}$ below $0.7$ correspond to temperatures below $15$ K, and while it is expected to find very few systems that exhibit this behaviour, we have decided not to limit the range of $U_{\rm min}$ and allow it to vary between $0.1$ and $50$, to capture even the most extreme cases. It has been shown that at least in the case of local galaxies in the Spitzer Nearby Galaxy Survey (SINGS), the $U_{\rm min}$ can be limited between $0.7\leq U_{\rm min}\leq25$, and a MW - like dust model can be adopted to limit $q_{\rm PAH}$ between $0.004$ and $0.046$. Incorporating the use of the optimised set of parameters, which has been done in similar studies (e.g., \citealt{magdis2012}; \citealt{magnelli2012}; \citealt{santini2014}), might have a positive effect on computational speeds, however we need to consider that this reduced parameter space has only been robustly verified for nearby solar-like metallicity populations of galaxies and might otherwise risk under(over)estimating the dust masses for extremely cold(warm) systems.\par
We can thus extract the following physical parameters from the fit - $\gamma$, 
$q_{\rm PAH}$, $U_{\rm min}$. The dust mass is simply computed from the normalisation, while the $L_{\rm IR}$ is $L_\nu$ integrated over the 8 - 1000 $\mu$m range.
As an additional parameter, we can obtain $\langle U \rangle$ by either utilising the \autoref{eq:5}, or alternatively, as prescribed by \citet{magdis2012}, we can use \autoref{eq:4}, where we set $L_{\rm dust}=L_{\rm IR}$ and $P_{0}\approx 125$.

\subsection{Removing AGN contamination}
Active galactic nuclei (AGN) can have a significant impact on the ISM of galaxies that host them. They possess an ability to halt star-formation by heating up the gas and dust or completely quenching the galaxy by stripping away its fuel. Under the common assumption \citep{antonucci1993,urry1995,tristram2007}, AGN are surrounded by dusty tori, that similarly to the ISM dust, can absorb the UV/optical light from the AGN and re-radiate it at redder wavelengths, normally peaking in the mid-infrared (MIR) regime at ~20-50\,$\mu$m. It has also been shown that for select extreme cases \citep{Mullaney2011}, the AGN emission dominates the SED of a galaxy even at $60\,\mu$m, which presents a new challenge when calculating an infrared luminosity of a source. Infrared derived star-formation rate estimates rely on a robust understanding of the \lir. Therefore, it is imperative to separate the energy contributions from hot dust in the ISM and a possible AGN. \par

In order to account for the effects of the IR contamination by AGN when calculating $L_{\rm IR}$, as well as to identify all the possible systems that might contain an active nucleus in our sample, we have decided to adopt a set of AGN templates from \citet{Mullaney2011} (M11). These templates have been empirically derived by assuming a modified blackbody function and fitting it to a set of Swift-BAT AGNs as well as IRAS spectra. The obtained models describe intrinsic AGN emissions in the range spanning from MIR to FIR (6-100\,$\mu$m). In this case, a typical AGN SED  could be thought of as a broken power law at $\leq 40 \mu m$, that rapidly vanishes when moving above $40$ $\mu$m. The average intrinsic AGN emission can be described as follows:
\begin{equation}
F_\nu =
 \begin{cases}
    \lambda^{1.8}       & \quad \text{at} \quad 6 \, \mu  \text{m}<\lambda<19\,\mu \text{m} \\
    \lambda^{\alpha}    & \quad \text{at} \quad 19 \,\mu  \text{m}<\lambda<40\,\mu \text{m} \\
    \nu^{1.5}F_\nu^{BB} & \quad \text{at} \quad \lambda>40 \, \mu \text{m}
 \end{cases}
\end{equation}
where $F{_\nu}^{BB}$ is the modified blackbody function and $\alpha$ is the spectral index. In our procedure we utilise the high and the low luminosity templates, with the $\alpha=0.0$ and $0.4$ respectively. In addition to that, we allow for a linear combination between the two, with varying coefficients, thus expanding the existing template space.

\subsection{Stellar emission component}
The contribution of stellar emission in the observed NIR bands, such as $Spitzer$ IRAC 1-4, can be quite significant, especially when we move to higher $z$. Therefore, if no libraries representing the luminosity from dust attenuated stellar light are available, one might either underestimate the slope of the AGN or overestimate its normalisation. This can lead to erroneously assigning more luminosity to the
AGN component in the MIR and therefore underestimating the \lir. \par
To avoid this, we additionally incorporate a library of stellar emission models, which are an updated version of the templates described by \citet{brammer08} (GB08). These are based on the stellar population synthesis models \citep{conroy09}, that were optimised for deep optical-NIR broadband surveys. The models form a basis set of a larger library, and were derived by using the 'non-negative matrix factorization' algorithm that was described in \citet{br07}. The method attempts to reproduce the full library of templates, by finding a non-negative linear combination of a smaller number of models. These can be considered as the 'principal component' blueprint of the larger catalogue. In total there are $12$ optical SEDs, which include both dust attenuated and non-attenuated starlight. We have incorporated these models into our fitting routine, and if the UV-optical photometry is available, this allows us to extract properties such as $M_{*}$, SFR, $E(B-V)$ and SFH. If no UV-optical data is available, the addition of these templates is still useful, as they can account for the excess flux in the rest-frame NIR bands, in conjunction with AGN and IR templates.
\par
\begin{figure}
\begin{center}
\includegraphics[width=0.52\textwidth]{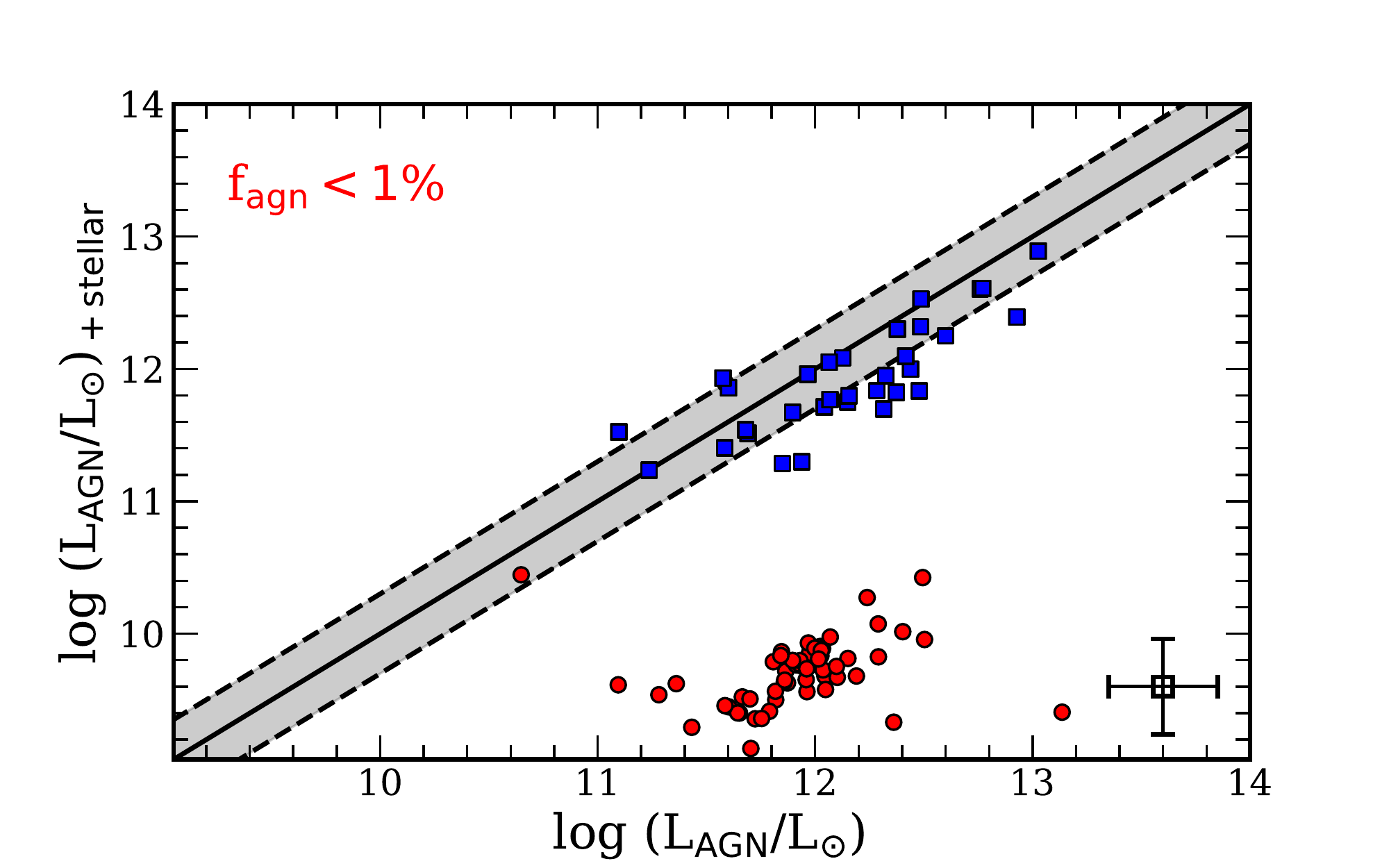}
\caption{A comparison between fits that incorporate the stellar component, and those that do not for $z>2$ galaxies. Blue squares and red circles represent objects which were assigned $f_{\rm AGN}>0.01$ and $f_{\rm AGN}<0.01$ respectively, by the means of our three component fit. The median uncertainty for both quantities is in the bottom right corner. The  black  solid  line  represents  the  1:1  relation  and  the  grey  regions  cover  the  0.3  dex  range.}
\label{fig:stelvsnostel}
\end{center}
\end{figure}

To test how our fits behave without the stellar component, we have isolated $\sim 100$ objects with $z > 2$, so that our bluest available band traces $\lambda_{\rm rest}\sim1\,\mu$m, where the contribution from stellar emission becomes non-negligible. We then exclude GB08 templates and re-fit our objects. This results in two outcomes that can be seen in \autoref{fig:stelvsnostel}. We find that, by removing the additional component, we tend to either overestimate the AGN contribution (blue squares), with the median being $\sim 1.5$, or alternatively erroneously assign a galaxy to contain an AGN (red points).

\subsection{Bringing it all together}
\begin{figure}
\begin{center}
\includegraphics[width=\columnwidth]{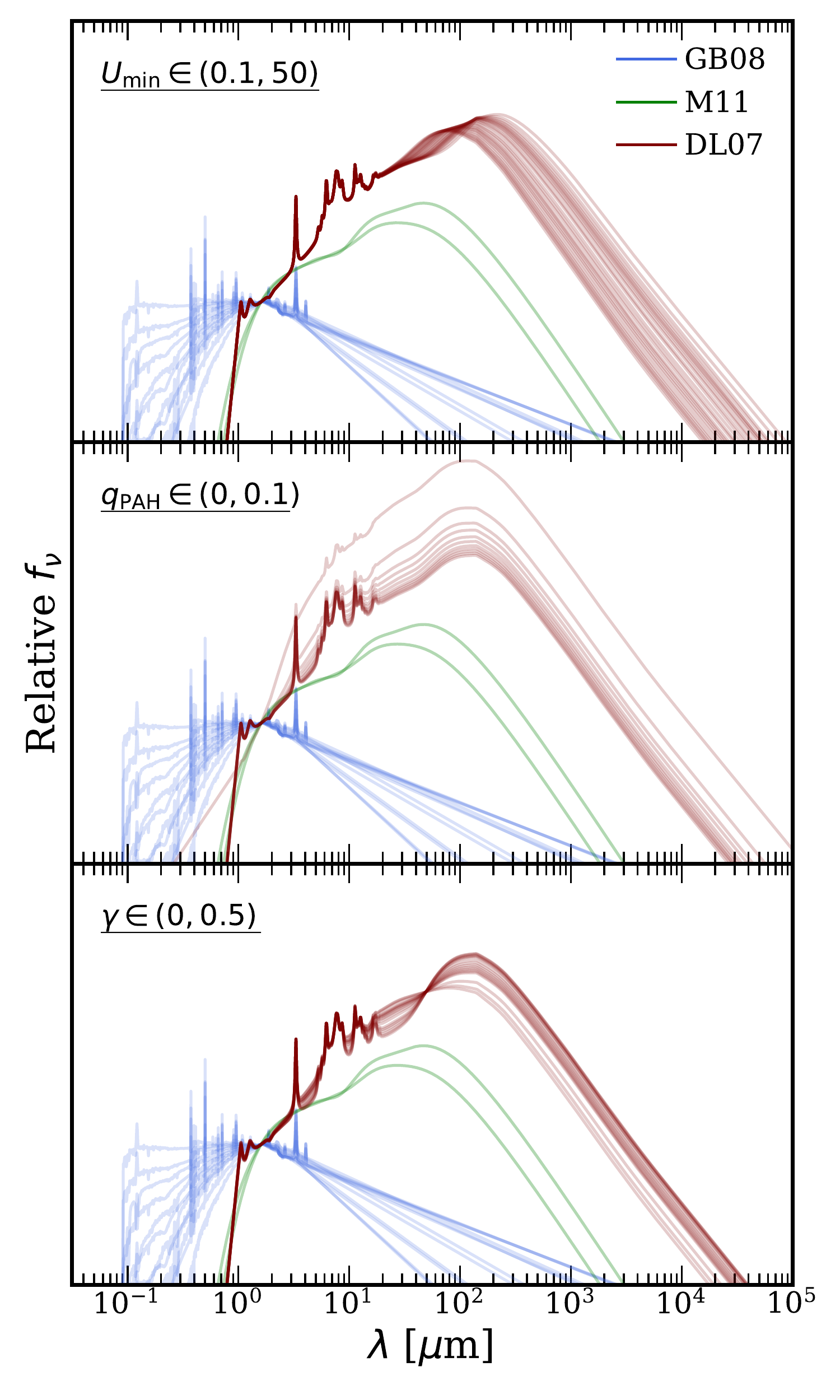}
\caption{A compilation of all template flavours used in \texttt{Stardust}. Colour-coding is blue, green and maroon for \citealt{brammer08} UV-Optical SPS models, \citealt{Mullaney2011} AGN models, and \citealt{draine07} IR dust models respectively. We show the variations in $U_{\rm min}$, $q_{\rm PAH}$, and $\gamma$ separately on each panel, while other parameters are fixed. For visualisation purposes all templates shown here have been normalised to the $K$-band.}
\label{fig:temps}
\end{center}
\end{figure}

In order to model an SED of a galaxy and extract the physical parameters, we first transport all three components - the stellar, the dusty torus AGN and the infrared dust emission to a common wavelength grid spanning the range from 10$^{-4}$ to 10$^5\,\mu$m. Our method relies on a linear combination of these models, thus making it imperative for them to share a common range, so that co-adding them is made possible.
In certain cases where this grid falls outside the original range of the template, we extrapolate bluewards and redwards by using a steeply declining power law. This was done to ensure that the resultant galaxy emission is continuous without any sudden breaks, which could interfere with the fitting, where one of the templates has ran out of range. These added power laws do not introduce any additional emission, as the flux density contribution from them is orders of magnitude lower than that of the original template. We then redshift the wavelength grid on a per galaxy basis and normalise the templates to ensure that they are not separated by tens of orders of magnitude. In \autoref{fig:temps} we show all the templates used in \texttt{Stardust}, normalised to the $K$-band, and in \autoref{tab:template-params} we list all the relevant parameters of the models.  \par

\begin{table}
\caption{Template parameters used for \texttt{Stardust} fit.}
\centering
\footnotesize
\label{tab:template-params}

\tabcolsep=0.1mm
\begin{tabular}{cc}
\hline\hline
Parameter & Value \\
\hline
\multicolumn{2}{c}{Dust emission: DL07 (Updated in \citealt{draine14})}\\
\hline
\hline
& $[0.1,0.2,0.3,0.4,0.5,0.6,0.7,0.8,$\\
$U_{\rm min}$ & $1.0,1.7,2.0,3.0,4.0,5.0,7.0,8.0, $\\
 & $10.0,12.0,15.0,20.0,25.0]$, 26 values \\
\hline
$q_{\rm PAH}$ & $[0; 0.1]$, 11 values in steps of 0.1\\
\hline
& $[0,0.001,0.0025,0.005,0.0075]$\\
 $\gamma$ & $+[0.01; 0.1]$, 9 values in steps of 0.01\\
 & $+[0.2,0.35,0.5]$, 17 values in total\\
\hline
\hline
\multicolumn{2}{c}{Optical emission: \citealt{brammer08}$^a$}\\
\hline
$A_V$ \citep{calzetti00} & [0.6,0.12,0.19,0.29,1.05,2.68,\\
& 0.11,0.36,0.98,1.54,1.97,2.96]\\
\hline
$M/L_V$ & [0.38,0.76,1.68,4.01,6.45,44.48,\\
& 0.12,0.21,0.33,0.64,1.57,4.00]\\
\hline
log$_{10}$(sSFR) & [-10.75,-11.37,-11.90,-12.53,-12.05,\\
& -12.47,-8.37,-8.60,-8.50,-8.57,-8.93,-8.90]\\
\hline
\hline
\end{tabular}
\begin{tablenotes}
\item[a] \footnotesize{Please refer to \citealt{brammer08} for a more detailed description of the creation and selection of these basis set templates. See \citealt{br07} for a methodology regarding the SFH.}
\end{tablenotes}
\end{table}

Subsequently, we perform a synthetic photometry on all three components separately in all observed bands where data is available. This is done by convolution of the filter transmission curves with the model SEDs. The resultant synthetic fluxes for each template and available observed band are then all combined into a two-dimensional matrix and passed onto the non-negative least square (nnls) algorithm in the python \texttt{scipy.optimize.nnls} package, that finds the best solution vector for that object. The nnls is a simple minimisation algorithm that in our case takes the form of,
\begin{equation}
\label{eq:nnls}
\text{min }||A_{b,t}x^{t}-y^{b}||_2  \quad \text{where} \quad x\geq0,
\end{equation}
with $A$ being the uncertainty weighted template matrix,  $y$ as the signal to noise for each band respectively, $x$ the solution vector and $||.||_{2}$ signifying the Eucledian norm. Instead of finding the best fitting template, the algorithm computes the best-fit coefficients $x$, with $x\geq0$.
The number of simultaneously fit templates is user defined. For our purposes we chose to fit all 12
GB08 and 2 M11 templates at the same time, each time combining them with a single DL07 template, and iterating through all the possible combinations, and then finding the best fit. Following the \autoref{eq:nnls}, this is done by building the matrix $A$, for each band $b$, and each template $t$, where the first 14 templates in the matrix remain fixed, while the last element is being continuously replaced with a new DL07 model and looped over. In the end we obtain $4,862$ possible best-fit vectors - $x^t$, one per each DL07 model. The final result is then extracted from the $\chi^{2}$ distribution of all best-fit solutions. The advantage of this method is in avoiding progressively looping over all possible template combinations to find the best solution, and instead only choosing to loop through DL07 templates. This approach significantly reduces the amount of required computational resources. The resultant solution vector encodes the individual contributions from each template to the total emitted flux in each band. These are then added together to return the best fit solution. These three component split allows us to predict exactly how much each component contributes to the source's \lir, thus allowing us to differentiate between the AGN and the warm ISM dust emissions. In addition to that, the normalisation of DL07 templates leads us directly to the number of hydrogen nucleons, linking it to dust mass via \autoref{eq:4}. The radio data points are not being considered by our fitting routine, however we add a power-law radio slope with a spectral index of $-0.75$, as described by the FIR-radio correlation in \citet{delv20}, mainly for visualisation purposes, but also to detect the existence of the AGN radio excess.

\subsection{Calculating the IR properties} \label{sec:da}
To derive luminosity estimates, we integrate the three component summed SED in the $8-1000\,\mu$m. This gives us the $L_{\rm IR,tot}$ that contains within itself the energy emitted by the ISM dust and AGN torus, if present. The contribution of stellar emission at $\lambda_{\rm rest} > 8\,\mu$m is negligible, therefore we do not go through an additional step of subtracting those models. We then integrate the best fit template with just the AGN contribution in it, to obtain the $L_{\rm AGN}$ and $L_{\rm IR}=L_{\rm IR,tot}-L_{\rm AGN}$. We also compute the $f_{\rm AGN}=L_{\rm AGN}/L_{\rm IR,tot}$ to estimate how strongly the infrared SED of a galaxy is contaminated by AGN activity. In addition to that, it allows us to separate our sample into objects that have an active nucleus and those that do not. The conditions of the ISM in these different environments may vary quite significantly and would affect the extracted scaling relations if not treated correctly.\par
The normalisation of the DL07 models returns the number of hydrogen nucleons as one of its free parameters, from which we can obtain the fiducial $M_{\rm H}$. We then compute \md\ by converting this quantity assuming a fixed gas-to-dust ratio of $\delta_{GDR}=100$, as prescribed in DL07. It is important to note that this ratio is encoded into the models and does not represent an actual physically meaningful conversion factor.
We also compute a $T_{\rm dust}$ proxy in the form of the average radiation field intensity $\langle U \rangle$, from \autoref{eq:4}, by assuming a $P_{0}=125$. We note however that this quantity represents the luminosity-weighted dust temperature and has little to no bearing on the temperature of the cold dust or gas (see e.g.\ \citealt{scov2016,scov17}).

\section{Stellar Mass Comparison} \label{sec:appendix-c}
To test the robustness of  \texttt{Stardust} - derived $M_*$ estimates we start by first comparing them to the $M_*$ as given by the original catalogue. The SDC2 takes its $M_*$ directly from the COSMOS 2015 catalogue \citep{laigle2016}, which uses SED fitting code \texttt{LePhare} \citep{lephare1,lephare2} to constrain both the photometric redshift, $M_*$, as well as a host of other parameters. Similarly to our approach, \texttt{LePhare} relies on \citet{ssp} SSP libraries to fit galaxy SEDs. We begin our fitting procedure by carefully correcting all the COSMOS 2015 aperture fluxes to total fluxes, as well as correcting for the MW extinction, as prescribed in \citet{laigle2016}. We then cross-match these sources to SDC2, by using the $K$-band, and fit the entirety of available 36 bands with our code. As a sanity check, we additionally run the same photometric catalogue through \texttt{EAZY}, albeit stopping at IRAC 2.\par
\begin{figure*}
\begin{center}
\includegraphics[width=.8\textwidth]{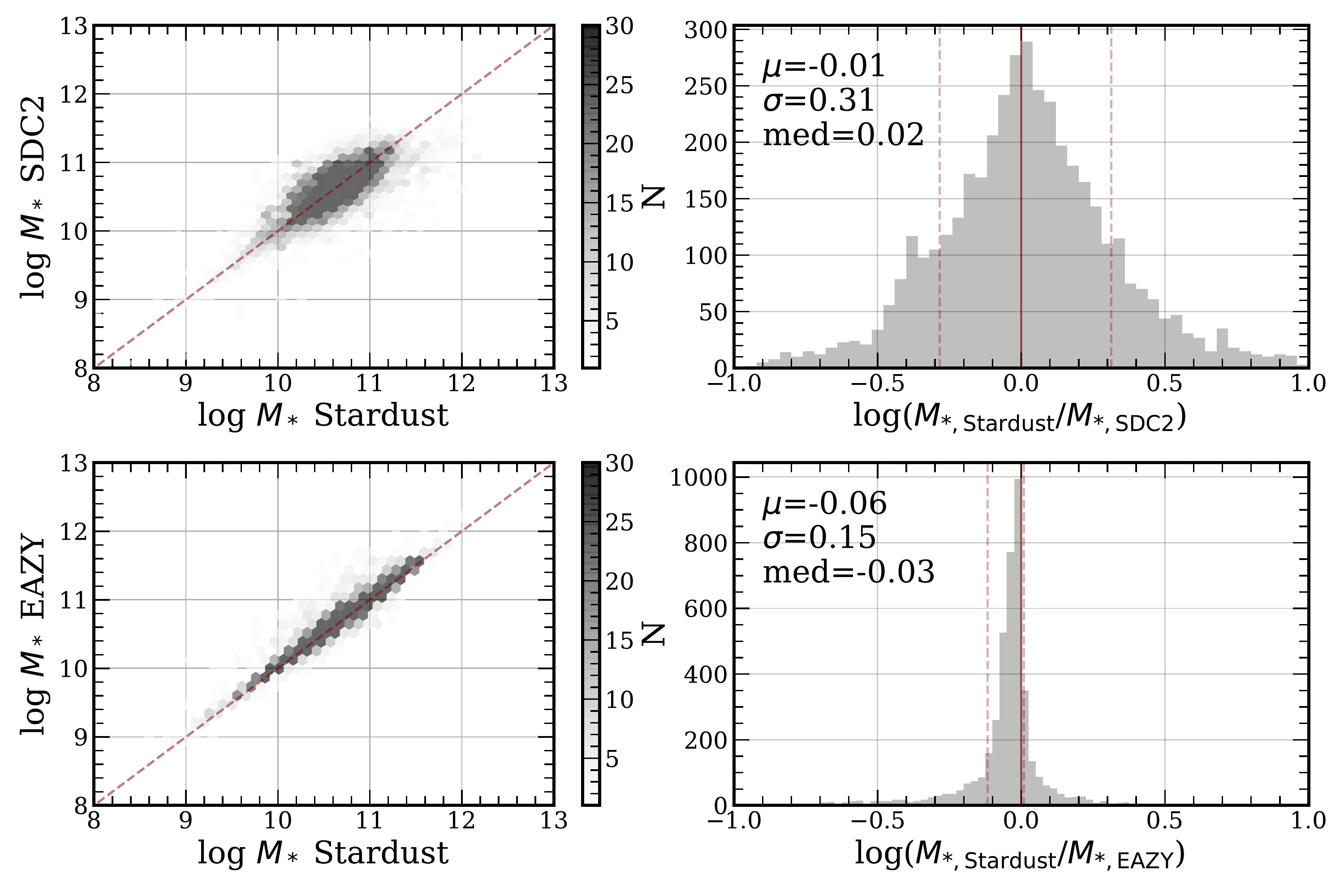}
\caption{Comparison of the \texttt{Stardust} derived $M_{*}$ vs SDC2 (top) and \texttt{EAZY} (bottom). The dashed maroon line represents a 1:1 relation. The solid and dashed maroon lines represent a 1:1 relation and the 68 \% confidence interval, respectively.}
\label{fig:mstar}
\end{center}
\end{figure*}

We present this comparison in \autoref{fig:mstar}. The $M_*$ given by the parent catalogue and the ones derived by \texttt{Stardust} derived $M_*$, agree very well, with \texttt{Stardust} on average under-predicting the $M_*$ by $\sim 0.01$ dex. We attribute a considerable 0.31 dex scatter to the fact that in their \texttt{LePhare} fit, \citeauthor{laigle2016} have used an iterative procedure, that involved correcting the observed fluxes in order to match the colours of the model library.
\par
More reassuringly, we find a good correlation when comparing our $M_*$ to \texttt{EAZY}, with a median offset of $-0.03$ dex and a minor 0.15 dex scatter being most likely induced by the fact that with \texttt{Stardust} we fit the entire available spectrum from UV to FIR, as opposed to just UV-optical, with \texttt{EAZY}.

\section{Comparison with \texttt{CIGALE}} \label{sec:appendix-d}
In order to better understand how our independent linear combination approach compares to the energy balance method, we have fit our sources with \texttt{CIGALE}. For this we have utilised a set of simple stellar population (SSP) models from \citet{ssp} for the non-obscured stellar light, the revised version of the \citet{dl07} templates for the obscured stellar light, reprocessed by dust, and \citet{fritz06} for the AGN contribution. The attenuation law that we considered was described in \citet{calzetti00}. For the SSP templates, we have assumed a single delayed SFH. These \texttt{CIGALE} fits should be treated as a `basic' first-pass approach, due to the computational limitations necessitating a constrained range for the template parameters. Ideally, such an analysis would require a flexible SFH, in order to obtain better SFR, as well as a wider range of parameters, both for the DL07 and \citeauthor{fritz06} templates. \par
\begin{figure*}
\begin{center}
\includegraphics[width=.8\textwidth]{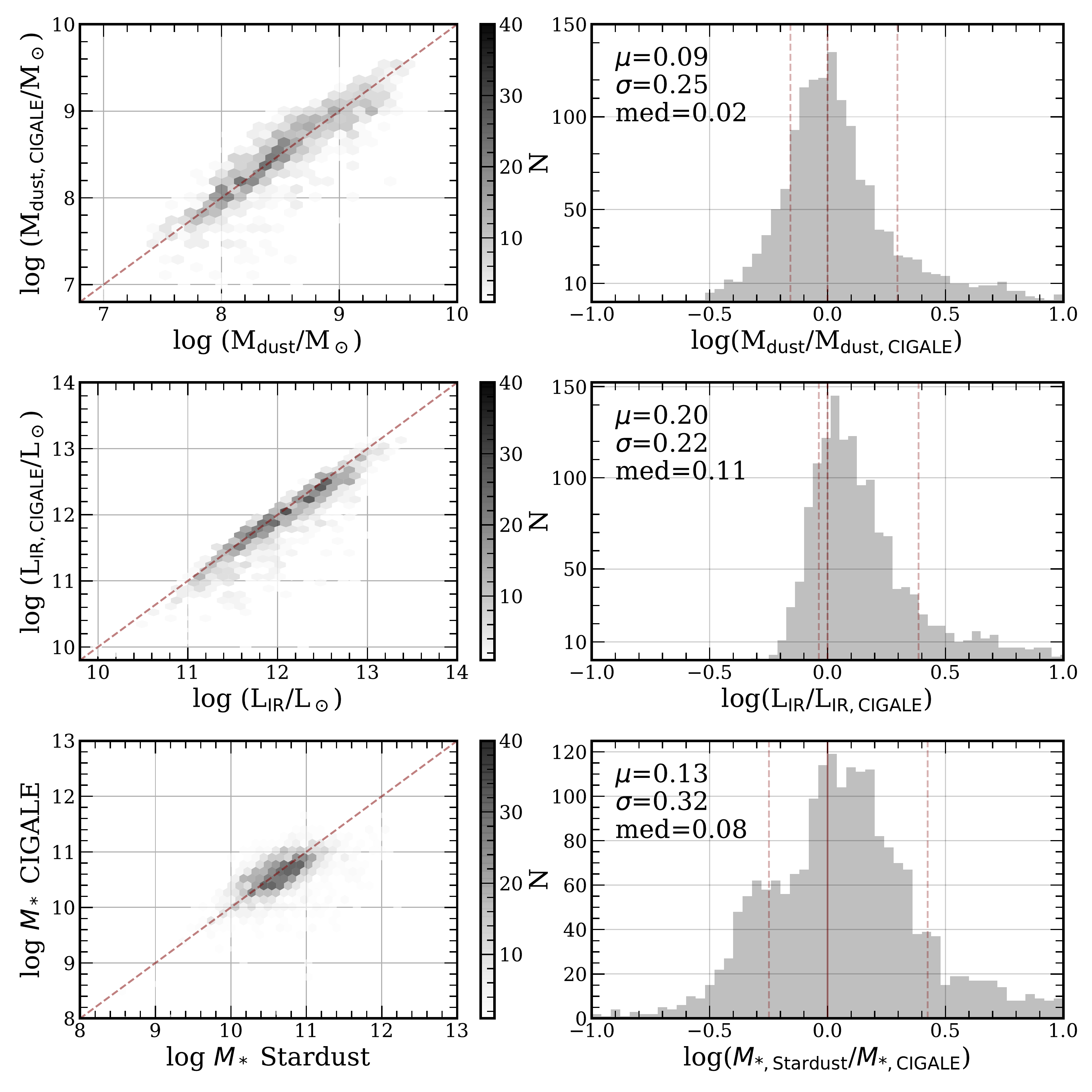}
\caption{Comparison of the derived $M_{\rm dust}$ (top) and $L_{\rm IR}$ (bottom) between \texttt{Stardust} and CIGALE. The solid and dashed maroon lines represent a 1:1 relation and the 68 \% confidence interval, respectively.}
\label{fig:cigale}
\end{center}
\end{figure*}

We show the comparison between the two methods in \autoref{fig:cigale}. There is a very good agreement between the $M_{\rm dust}$ derived with our code and \texttt{CIGALE}, with the difference having a mean of 0.09 dex and median of 0.02 dex. The derived values of $L_{\rm IR}$ are however in tension, with a mean of 0.20 dex and median of 0.11 dex. We attribute the significant outliers ($>1\sigma$) to cases where the energy balance method in \texttt{CIGALE} has failed to account for the extra FIR flux. We also compare the \texttt{Stardust} and \texttt{CIGALE} computed $M_*$, and find that the two agree within 0.1 dex, albeit with a significant 0.3 dex scatter. As we have already discussed in \autoref{sec:sed}, in certain environments the stellar and the dust emission could be spatially disconnected, thus the energy balance might not be the best physically motivated option. In addition, when dealing with extreme sources, the \citeauthor{calzetti00} attenuation law might not allow the energy balance approach to account for all IR flux (e.g. see \citealt{buat19}). The above, however, are not the only explanations, as the identification/matching problems as well as IR flux extractions could also play a part in creating this tension between our results and \texttt{CIGALE} (e.g. see \citealt{malek18}). \par
When directly comparing computation times, it is important to note that, \texttt{CIGALE} fits sources within redshift blocks, where it pre-compiles a set of models first and then estimates the best-fit parameters, while \texttt{Stardust} fits sources sequentially. As such, despite both methods being parallelised, it is difficult to achieve a fair comparison between the two. Within a single redshift block, that numbers 288 objects, \texttt{CIGALE} has computed $50\times10^6$ models, and found the best fit in about 2.5 hours. Due to how the linear combination is performed within \texttt{Stardust}, defining an exact number of models attempted is not possible. However, considering that the 12 optical templates have been constructed as a basis set of $\sim 3,000$ models described in \citet{brammer08}, combining that with 2 AGN templates and 
$\sim 4,800$ DL07 models, results in roughly $30\times10^6$ total effective model combinations. Our code then takes 14 minutes in total to fit the same 288 objects, which is approximately 11 times faster than \texttt{CIGALE}, for the same number of CPU cores.

\begin{deluxetable*}{ccc}
\tablecaption{\label{tab:catalogue} Structure of the best-fit catalogue$^{\rm a}$}
\tablehead{%
Column Name	& Units &  Description }
\startdata
ID & - & ID of the object from the parent catalogue\\
RA & deg & The right ascension coordinate, as given in the parent catalogue\\
DEC & deg & The declination, as given in the parent catalogue\\
Area & - & Same as \texttt{goodArea} flag in the parent catalogues\\
z & - & Redshift used for fitting \\
ztype & - & Redshift type, 1 for spectroscopic, 0 for photometric, and 2 for $z_{\rm phot}$ from \texttt{EAZY}\\
LIR$\_$total & L$_{\odot}$ & Total FIR luminosity, obtained as the sum between AGN and DL07 components\\
eLIR$\_$total & L$_{\odot}$ & Uncertainty on the total FIR luminosity\\
Lagn$\_$total & L$_{\odot}$ & AGN luminosity\\
eLagn$\_$total & L$_{\odot}$ & Uncertainty on the AGN luminosity\\
Lir$\_$draine & L$_{\odot}$ & Luminosity given by the DL07 template\\
eLir$\_$draine  & L$_{\odot}$ & Uncertainty on the luminosity given by the DL07 template\\
MD & M$_{\odot}$ & Dust mass as predicted by the best-fit DL07 template \\
eMD & M$_{\odot}$ & Uncertainty on the dust mass \\
deltaGDR & - & Gas-to-dust ratio \\
MG & M$_{\odot}$ & Gas mass computed from $\delta_{\rm GDR}$ and \md \\
eMG & M$_{\odot}$ & The uncertainty on gas mass \\
Mstar & M$_{\odot}$ & Stellar mass, equal to the one in the original catalogue\\ 
lastdet & $\mu$m & Last band that has a S/N$>3$ detection ($\lambda_{\rm last}$). Given in rest frame.\\
chi2 & - & The $\chi^2$ of the best-fit coefficients\\
f$\_$agn & - & AGN fraction, given as L$_{\rm AGN}$/L$_{\rm IR,total}$\\
efagn & - & Uncertainty on the AGN fraction\\
fgas &	- &	Gas fraction computed as $M_{\rm gas}/M_{\rm dust}$\\
fgas$\_$FMR &	- &	Gas fraction computed assuming $\delta_{\rm GDR}=100$\\
Umin & - & Best fit $U_{\rm min}$\\
gamma & - & Best fit $\gamma$\\
qpah & - & Index of the best fit $q_{\rm PAH}$ value\\
U & - & Average radiation field intensity \avu \\
sU & - & Uncertainty on \avu \\
deltaMS	&  - &	SFR/SFR$_{\rm MS}$ where SFR=$L_{\rm IR,draine}\times10^{-10}$ and SFR$_{\rm MS}$ is given by \citealt{schreiber15}\\
e$\_$deltaMS & - & Uncertainty on $\Delta$MS\\
\enddata
\begin{tablenotes}
\item[a] \footnotesize{\texttt{Stardust} also returns $M_*$, $A_{\rm V}$ and UV-optical SFR, these are not included in the release version of the catalogue, but are available upon request.}\\
\item[b] \footnotesize{We have used the 7970d55 (28/06/21) version of \texttt{Stardust} to produce these catalogues.}
\end{tablenotes}
\end{deluxetable*}
\clearpage
\bibliographystyle{aasjournal}
\bibliography{refs}



\end{document}